\documentclass{aa}  
\usepackage{graphicx}
\usepackage{graphics}
\usepackage[flushleft]{threeparttable}
\usepackage{txfonts}
\def\hbeta{H$\beta$}

\def\lsigma{$L-\sigma$}

\def\eqw{\rm EW(H$\beta$)}
\def\lbeta{L(H$\beta$)~}
\def\lhbeta{L(H$\beta$)~}
\def\lsigma{$L-\sigma$~}

\usepackage[normalem]{ulem}
 
\def\msun{M$_{\odot}$}
\def\myou{$M_{young}$}

\def\kmsecmeg{\thinspace\kmsec\Mpc$^{-1}$}
\def\ergsqcmsec{\thinspace\hbox{erg}\sqcm\sec$^{-1}$}
\def\Zsun{\thinspace\hbox{$\hbox{Z}_{\odot}$}}
\def\Mpc{\thinspace\hbox{Mpc}}

\def\kmsec{\thinspace\hbox{$\hbox{km}\thinspace\hbox{s}^{-1}$}}

\def\sqcm{\thinspace\hbox{$\hbox{cm}^{2}$}}
\def\sec{\thinspace\hbox{s}}

\begin{document}

\title{The Stellar Populations of HII galaxies:}
\subtitle{A tale of three bursts}

\author{Eduardo Telles\inst{1} \and Jorge Melnick \inst{1,2} }
\institute{ Observat\'orio Nacional, Rua Jos\'e Cristino, 77, Rio de Janeiro, RJ, 20921-400, Brasil
  \and
  European Southern Observatory, Av. Alonso de Cordova 3107, Santiago, Chile}

\offprints{Eduardo Telles\email{etelles@on.br}}
\date{}

\authorrunning{Telles \& Melnick}

\titlerunning{The Stellar Populations of HII galaxies}

\label{firstpage}

\abstract
    {}
   % aims heading (mandatory)   
    {We present a  UV to mid-IR spectral energy distribution study of a large sample of SDSS DR13 HII galaxies. These are selected as starburst (EW(H$\alpha) > 50$\AA) and for their high excitation locus in the upper-left region of the BPT diagram. Their photometry was derived from the cross-matched GALEX, SDSS, UKDISS and WISE catalogues.}
    % methods heading (mandatory)  
{We have used CIGALE modelling and SED fitting routine with the parametrization of a three burst star formation history, and a comprehensive analysis of all other model parameters. We have been able to estimate the contribution of the underlying old stellar population to the observed equivalent width of H$\beta$ and allow for more accurate burst age determination.}%%%%%%%%
  % results heading (mandatory)
{We found that the star formation histories of HII Galaxies can be reproduced remarkably well by three major eras of star formation. In addition, the SED fitting results indicate that: i) in all cases the current burst produces less than a few percent of the total stellar mass: the bulk of stellar mass in HII galaxies have been produced by the past episodes of star formation; ii) at a given age the H$\beta$ luminosity %(L(H$\beta$)) 
depends only on the mass of young stars favouring a universal IMF for massive stars; iii) the current star formation episodes are {\it maximal} starbursts, producing stars at the highest possible rate.}%%%%
{}

 
\maketitle

\section{Introduction}
\label{intro}

HII Galaxies are compact dwarf starburst galaxies with strong and narrow emission lines superposed on a weak blue continuum.  The optical spectra of HII Galaxies are indistinguishable from those of Giant HII regions in local galaxies \citep{SS1970}.  It is now widely accepted that they are not bona fide young galaxies forming their first generation of stars, as thought in the past, since they all show a population of old stars.
%As in fact, all galaxies seem to have old stars implying a commom birth date or an era in the history of the universe in which stars first formed within galaxies \citep[and references therein]{conse2014}.

\cite{Westera2004} used the spectra of some 100 HII galaxies to assess their stellar population content and history by deriving  absorption line indexes (based on H$\delta$, H+K(Ca), Mg$_b$ and D4000) and comparing with stellar population models of SB99 \citep{sb99} and BC03 \citep{bc03}.
%with the BaSeL stellar spectral library \citep{Westera2002}.
The main conclusion from that work is that, mostly, we can parametrize the star formation history (SFH) of HII galaxies with these three main stellar populations.
%\citep[see e.g.][W04]{Westera2004}.

Optical \citep{Telles1997a,Telles1997b} and near-IR imaging \citep[and references therein]{Lagos2011} have also convincingly shown that these dwarf starburst galaxies possess underlying old populations. Simulations also show the episodic nature of star forming galaxies, particularly at low masses \citep[see e.g][]{Pelupessy2004, Debsarma2016}, being three episodes the simplest choice in this scenario.

The morphologies of HII galaxies  remain, as first described by \cite{loose1986,kunth1988}, a mixed bag. The general properties of HII galaxies and Blue Compact Galaxies (BCG) broadly overlap \citep{kunthostl2000}. They have irregular shapes, typically small physical sizes, no signs of ordered structures, such as disks. Their starburst regions, consisting of emsembles of massive ionizing clusters and their respective giant HII regions, cover most of the extension of their optical images. More luminous HII galaxies seem to show some evidence of tails, fuzz in their outermost isophotes, and more disturbed overall morphologies whereas the lower luminosity ones seem more compact \citep{Telles1997a}. Deeper optical imaging \citep{Lagos2007} and near-IR imaging \citep{Lagos2011} reveal the  clumpy nature of their starburst regions.
  The various sub-classification attempts, such as cometary \citep{papa2008}, local tadpoles \citep{debra2012}, green-peas \citep{carda2009}, etc, all fall within the mixed bag of clumpy morphology with no fundamental differences in their intrinsic properties. In any case, due to their low mass, low oxygen abundance, low dust content, and low density environments,  HII galaxies constitute the simplest starbursts at galactic scales.

With the advent of large surveys, particularly the Sloan Digital Sky Survey \citep[][SDSS]{york2000}, star forming galaxies all fall back into a uniform spectroscopic class and are viewed in a more general common perspective.  Total stellar mass seems to be the main driver of the properties of the star forming galaxies at low redshift \citep{brinch2004, tremon2004}. However, the locus where emission line galaxies fall in the BPT diagram \citep{bpt1981}  determines some important general properties as well, since the star forming sequence is also a sequence of increasing excitation with the decrease of stellar mass and metallicity \citep[and references therein]{curti2017}.

HII Galaxies are particularly interesting as cosmological probes over a surprisingly large range of redshifts extending to redshifts of z=3-4 with present-day instrumentation. Their  emission-line luminosities, in particular the Balmer lines, correlate extremely well with the velocity-widths of the same emission lines \citep{ter81,mel88,telles93,bor09,bor11,chavez2014}. This so-called L-$\sigma$ relation can be calibrated as a distance indicator using Giant HII regions in local galaxies, and can thus be used to determine cosmological parameters \citep{mel2000,plionis11,ter15,chavez2016,Arenas2017}. Since the L-$\sigma$ method is independent from other methods, a cross comparison of results help us better understand the systematic uncertainties in these different methods, most notably the SNIa.

%%%%%%%%%%%%%55
While it seems clear that the L-$\sigma$ relation stems from the natural relation between the ionizing flux of a starburst and the mass of its ionizing cluster, the relation is empirical and thus suffers from considerable intrinsic scatter. In \cite{mel2017} we have explored ways of reducing the scatter, but stumbled against the difficulty of accurately measuring the ages of the young component. The traditional age indicator, the equivalent width of H$\beta$ (EW(H$\beta$)), is biased by contamination of the continuum by older underlying populations and therefore age corrections using EW(H$\beta$) tend to increase the scatter rather than reduce it.

In this paper  we make use of multi-wavelength stellar population
analysis by using the method of fitting the spectral energy
distribution (SED) from the UV to MIR in order to describe the
simplest star formation history for HII galaxies that accommodates
their general properties.  We wish to investigate how efficient HII
galaxies are in forming stars in the present burst as compared to
their past.  We also retrieve the true distribution of EW(H$\beta$) for
the young stellar component of our sample by applying a correction
factor $f_r$ that accounts for the contamination of the continuum by
the old stellar population derived from our stellar population
analysis.  This will help us further understand, and possibly reduce,
the systematic errors related to the use of the L-$\sigma$ as a
powerful indicator of cosmological distances.  Sec.~\ref{data}
presents our data selection of extreme star forming galaxies from the Sloan
Digital Sky Survey. Sec.~\ref{cigale} describes our SED fitting model, model choices, and procedure.  In
Sec.~\ref{results} we present our results, and conclusions are given
in Sec.~\ref{conclusions}.

%%%%%%%%%%%%%%%%%%%%%%%%%

 \section{Data and general spectral properties}\label{data}

 \begin{figure}
   \centering
\includegraphics[width=0.45\textwidth]{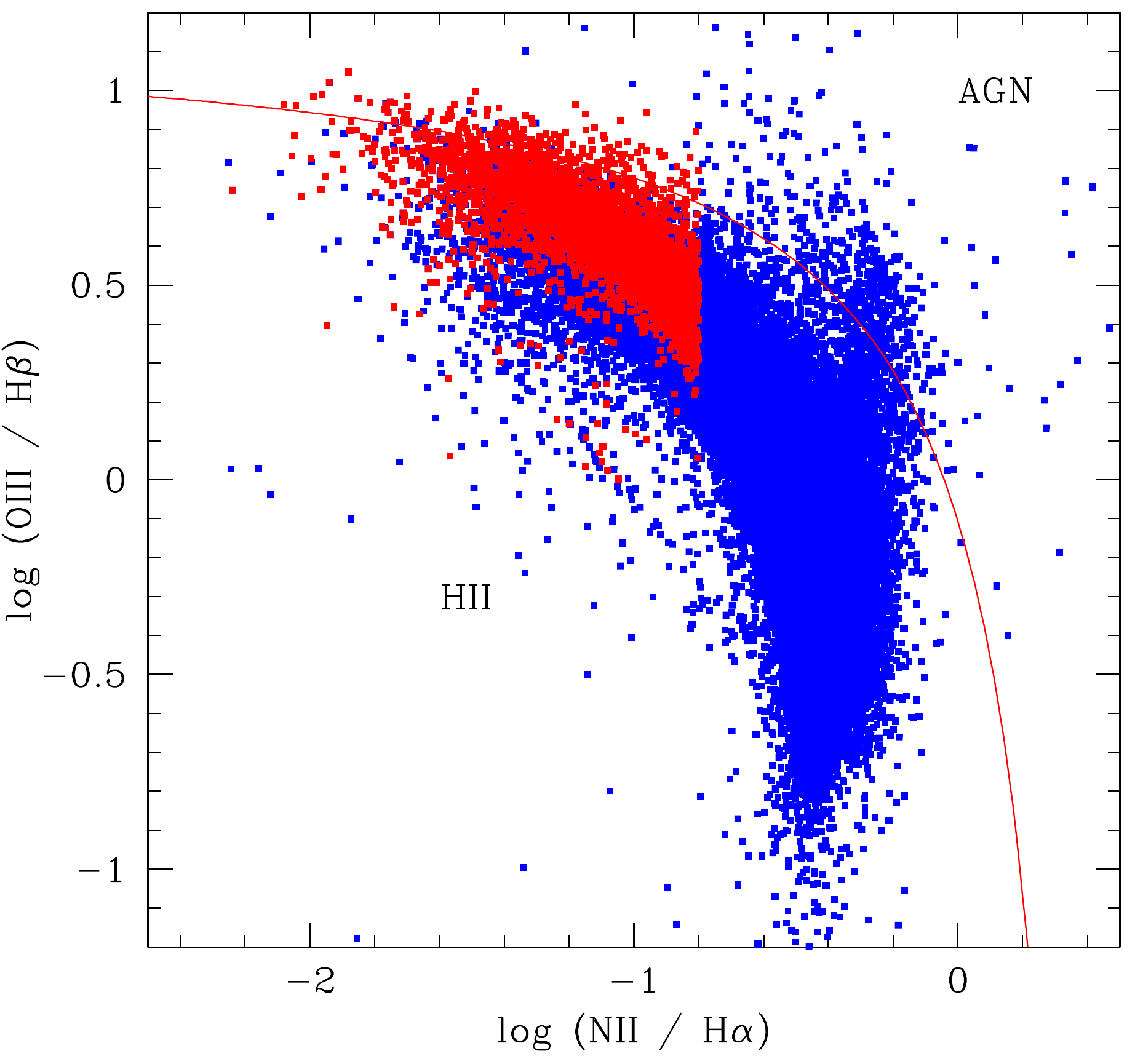} 
% \vspace*{-3.0cm}
\caption{\small BPT diagram of the selected objects. The blue points are the whole sample of 67000 starburst galaxies. The red points are the resulting spectroscopic sample of $\sim$4200 galaxies with the criteria given in Table~\ref{crit}. HII galaxies lie below and to the left of the  \cite{kauff2003} classification line (solid red line) that distinguishes AGNs from star forming galaxies.}
\label{one}
\end{figure}

 Our data are selected from the SDSS DR13 release \citep{dr13}
 cross-matched with the emissionLinePort table\footnote{the
   emissionLinePort table (Portsmouth stellar kinematics and
   emission-line flux measurements Thomas et al. (2013) are based on
   adaptations of the publicly available codes Penalized PiXel Fitting
   (pPXF, Cappellari \& Emsellem 2004) and Gas and Absorption Line
   Fitting code (GANDALF v1.5; Sarzi et al. 2006) to calculate stellar
   kinematics and to derive emission line properties. )}, and contains
 galaxies classified as subclass STARBURST which implies ${\rm EW(H}\alpha)>50$\AA.  These criteria reflect in over 67000 galaxies. From these
 we selected only those with ${\rm EW(H}\beta)>30$\AA\ and those whose
 line ratios fall within the upper left panel of the canonical
 interval for star forming regions in the BPT diagram \citep{bpt1981,kewley2006}.  These choices aim at selecting extreme star forming
 galaxies with high excitation, low abundances and low masses, typical
 of bona fide HII galaxies.  So, our selection criteria are
 more restrictive and do not include more luminous star forming
 galaxies. These criteria allow the inclusion of the lowest
 metallicity objects which have slightly lower [OIII]/H$\beta$ ratios
 due to their low ionic abundances \citep{izo2017}.  In order to avoid
 including local giant HII regions in nearby galaxies we also
 restricted by z > 0.005, resulting in $\sim$4200 SDSS objects.  Figure~\ref{one} shows the selected spectroscopic sample. A
 summary of these criteria is given in Table~\ref{crit}.

 \begin{table}
   \centering
\caption{Summary of the selection criteria of our spectroscopic sample, resulting in our SDSS sample of $\sim$4200 objects.}
  \begin{tabular}{c} \hline
EW(H$\alpha$) > 50\AA\\
EW(H$\beta$) > 30\AA\\
$0 < lg(\rm{[OIII]/H}\beta  < 1.2$ \\
$-2.5 < lg(\rm{[NII]/H}\alpha<-0.8$ \\
$ 0.005 < z < 0.4$ \\ \hline
\end{tabular}
\label{crit}

\end{table}

For our final sample of HII galaxies for our multiwavelength analysis from FUV to MIR, we chose from the SDSS sample only targets with GALEX  unique
photometry in both FUV (0.1528$\mu$m) \& NUV (0.2271$\mu$m) bands \citep[and references therein]{bianchi2014}. The resulting GALEX+SDSS sample
of HII galaxies consists of 2728 objects.
For this sample, we cross-matched targets with The UKIRT Infrared Deep Sky Surbey \citep[UKIDSS]{ukidss}  Y (1.036$\mu$m), J (1.250$\mu$m), H (1.644$\mu$m), K (2.149$\mu$m) near-IR bands, and with the Wide-Field Infrared Survey Explorer \citep[WISE]{wise} W1 (3.4$\mu$m), W2 (4.6$\mu$m), W3 (12$\mu$m), W4 (22$\mu$m) \mbox{mid-IR} bands.

 \begin{table}
   \centering
\caption{Number of objects that have the corresponding photometric data from the given surveys.}
  \begin{tabular}{lc} \hline
    GALEX+SDSS & 2728\\
    GALEX+SDSS+WISE & 2447 \\
    GALEX+SDSS+UKIDSS & 950 \\
    GALEX+SDSS+UKIDSS+WISE & 863 \\
\hline
  \end{tabular}
\label{photsamp}

\end{table}

 \begin{figure}
   \centering
  \includegraphics[width=0.4\textwidth]{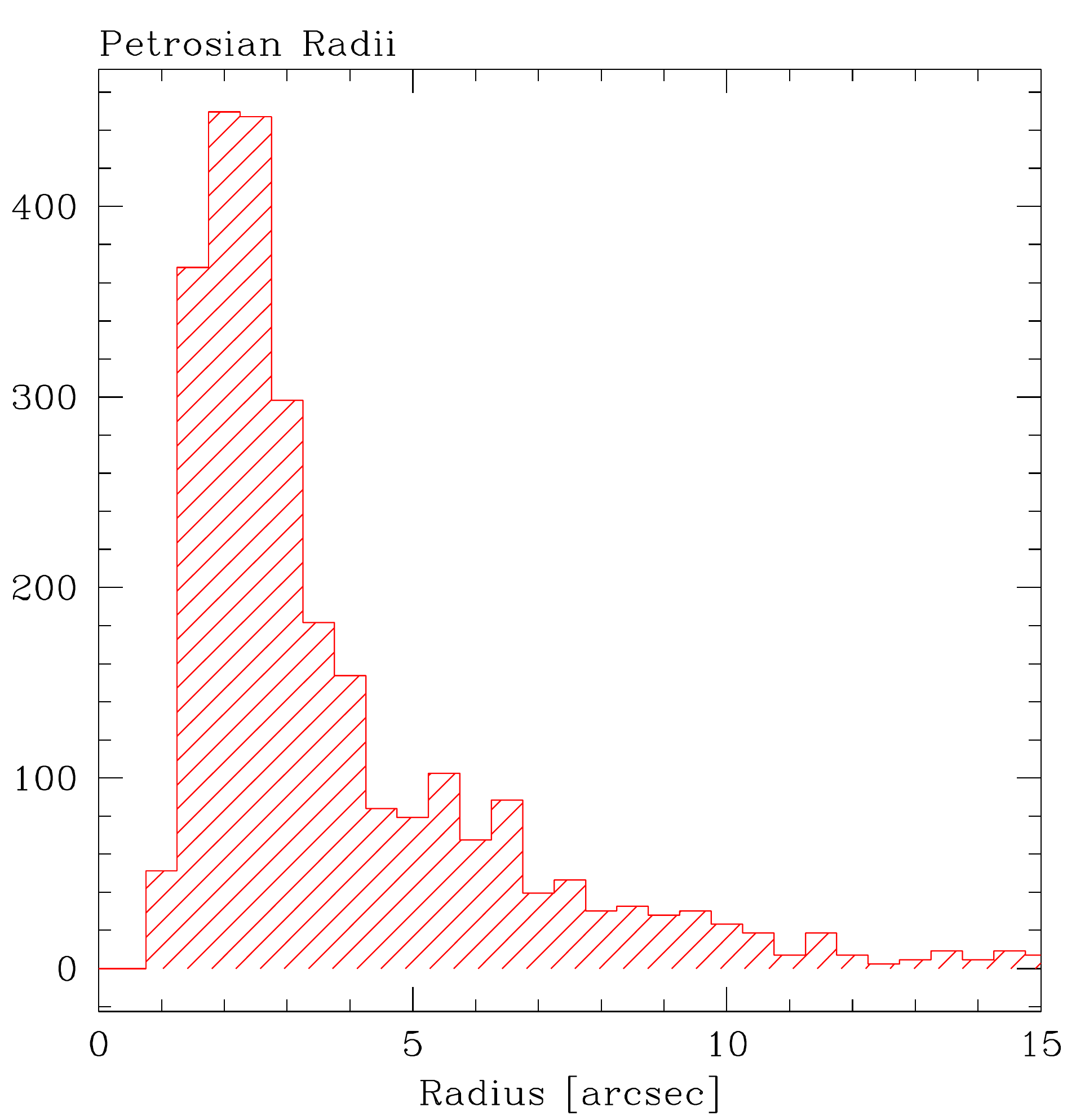}
\caption{\small Distribution of the Petrosian radius in the SDSS r band of our photometric sample of 2728 HII galaxies. The median value of the distribution is only 2.8\arcsec.}
\label{petro}
\end{figure}

In the end, we have chosen not to use the W3 \& W4 bands since we are most concerned with the stellar population properties as a result of our analysis, and these bands only suffer any significant effect at these longer wavelengths due to the dust emission component. Table~\ref{photsamp} shows the resulting photometric sample of HII galaxies.  The most restrictive photometric band is the near-infrared UKIDSS with NIR data for only one-third of our sample. In any case, in our analysis we use all available photometric data each individual galaxy. 
To minimize systematic effects we used Petrosian magnitudes except for GALEX for which we used model magnitudes.  Our choice of the Petrosian magnitudes ensures that in all bands we measured the fluxes in the same way to include the same percentage of the total flux. In any case, our objects are compact (see Figure~\ref{petro}), thus aperture effects are minimized since we are probably including all the flux in all bands.
  We also correlated with VISTA surveys but found no additional targets.

 \subsection{Basic properties}

 \begin{figure*}
   \centering
  \includegraphics[width=0.9\textwidth]{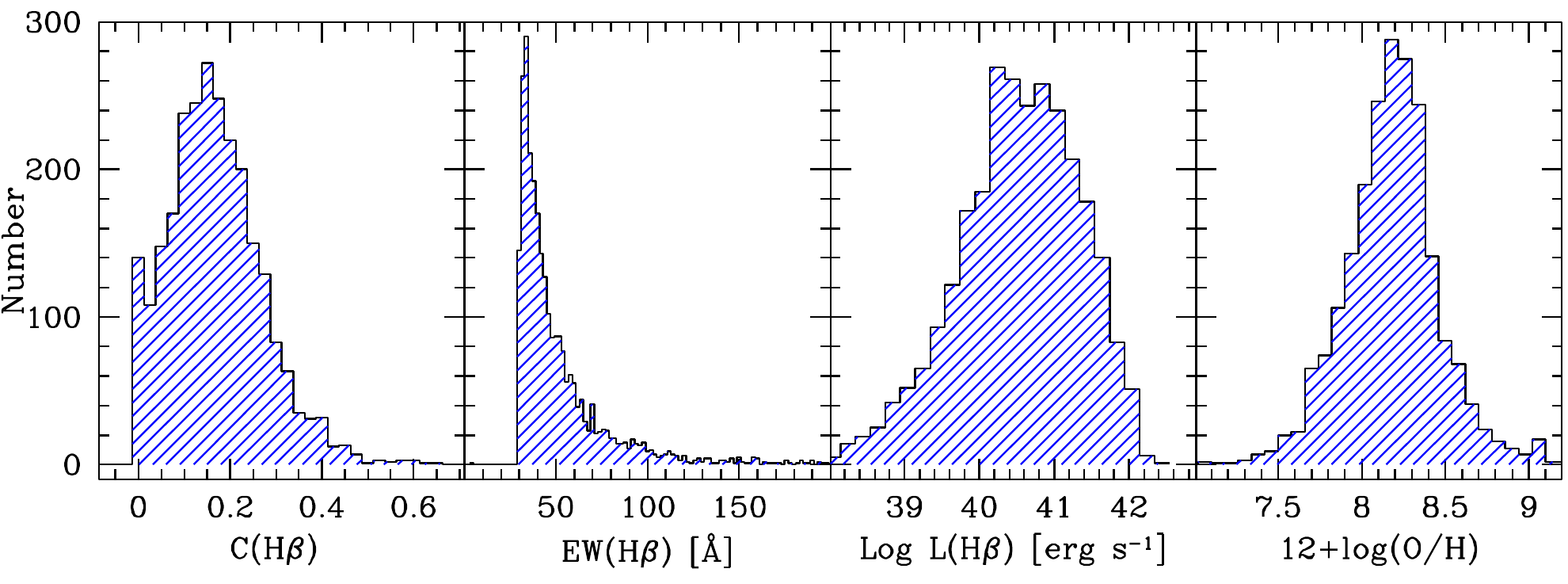}
  %  \includegraphics[width=0.24\textwidth,bb=39 158 545 635]{CHb_hist.pdf}
%  \includegraphics[width=0.24\textwidth,bb=39 158 545 635]{WHb_hist.pdf}
%  \includegraphics[width=0.24\textwidth,bb=39 158 545 635]{LHb_hist.pdf}
%  \includegraphics[width=0.24\textwidth,bb=39 158 545 635]{OH_hist.pdf} 
% \vspace*{-3.0cm}
\caption{\small Spectroscopic properties of our sample of intense emission line objects. (left) Distribution of the Balmer decrement derived extinction parameter. (middle left) Distribution of the equivalent width of H$\beta$. (middle right) Distribution of H$\beta$ luminosity. (right) Distribution of the oxygen abundance derived from the direct method for objects with the auroral [O III]$\lambda$4363\AA\  line detected.  This line is detected virtually in all our galaxies and allow for the determination of the electron temperature (see text).}
\label{sample}
\end{figure*}

 Figure~\ref{sample} shows the spectroscopic characterization of our  cross-matched GALEX+SDSS sample of 2728 HII galaxies. The Figure shows histograms of the logarithmic extinction correction factor C(H$\beta$), the observed equivalent width of H$\beta$, the derived H$\beta$ luminosity\footnote{Throughout this paper we assume H$_0$ = 71 \kmsecmeg.}, and the derived oxygen abundance (see below), respectively. A typical galaxy in our sample has low extinction (E(B-V) $\sim$ 0.1), intense emission lines (\mbox{EW(H$\beta$) $\sim$ 50 \AA}, $\log\rm{L(H}\beta)\sim40.5$ \ergsqcmsec), and low metal abundance (\mbox{$12+\log({\rm O/H}) \sim 8.2$}, Z$\sim 1/3$ \Zsun).

\subsubsection{Extinction corrections}

Our first step was to correct the fluxes for foreground extinction
using the maps of \cite{Schlafly2011} as reported in the SDSS data
base and the extinction law for the Galaxy from
\cite{cardelli1989}.  This is relevant because in these starburst
galaxies the foreground extinction is substantial, and the extinction
laws for the internal extinction are very different in the UV.

As a second step, the intrinsic internal reddening is then derived from the resulting H$\alpha$/H$\beta$ ratio (corrected for Galactic extinction) for each HII galaxy, using either a Calzetti  \citep{cal2000} or a Gordon \citep{gordon2003}  extinction law.
 The Gordon  extinction curve is that of the SMC bar which is the steeper curve in the UV. Their comparative results seem to indicate that there is a trend for the extinction curves to be steeper in the UV for systems of lower gas to dust ratios.  The starburst galaxies in our sample have typically sub-solar oxygen abundances implying a low dust content, and hence this ratio will be large for our sample galaxies, favoring a SMC-bar like extinction curve.  This agrees with previous findings by \cite{gordon1997} who found that the steeper UV extinction curve seems to be associated with enhanced star formation region such that of 30 Dor in LMC whose extinction curve differs from the rest of the LMC.

%We corrected the fluxes for foreground extinction using the maps of \cite{Schlafly2011} as reported in the SDSS data base and the extinction law for the Galaxy from \cite{Misselt1999}.

We then measured the line intensities for H$\alpha$, H$\beta$, and H$\gamma$ by optimizing the placement of the continuum and carefully adjusting the integration box to include all the line fluxes.
Inspection of the spectra showed that in general the higher Balmer lines ($\lambda\leq {\rm H}\delta4101$\AA) were embedded in a stellar absorption feature that appears to be significantly broader than the emission lines. For this reason we did not include these lines in our analysis. Even H$\gamma$ is in most cases somewhat affected by absorption although, considering that the equivalent widths of the absorption features in synthetic spectra for starburst are similar for H$\beta$ and H$\gamma$, we did not correct the line ratios for this effect.

%We finally introduced a further cut in the sample by demanding that the S/N in the continuum near H$\gamma$ be larger than 5 (S/N$>$5). This reduced the sample to 979 objects.

\begin{figure}
  \centering
  \includegraphics[width=0.45\textwidth]{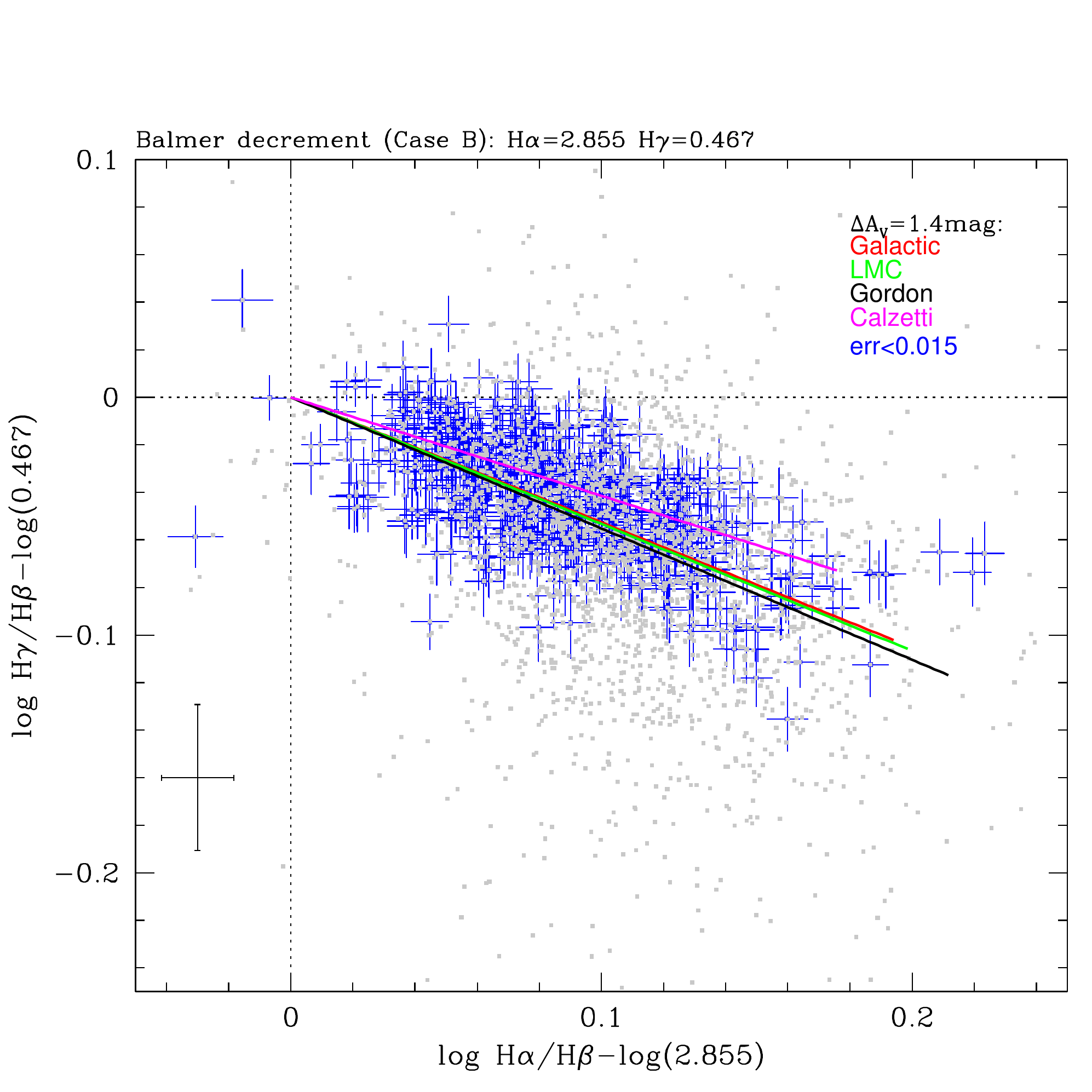} 
%\vspace*{-3.0cm}
\caption{\small Balmer decrements for our sample relative to the theoretical Case B recombination values for $T_e=11400$K.  The solid lines represent four different extinction laws as shown for a range of $\Delta A_V$ = 1.4 mag.  Points in gray are for the whole sample, while points in blue correspond to objects with errors in both axes $< 0.015$. The black cross on the left bottom represents the mean error of the whole sample.}
\label{two}
\end{figure}

In Figure~\ref{two} we plot the Balmer decrements divided by the theoretical (Case B) recombination values  for $T_e=11400$K, appropriate for the mean temperature of our sample, F(H$\alpha$)/F(H$\beta$)=2.855 and F(H$\gamma$)/F(H$\beta$)=0.467 \citep{Osterbrock1989}. Thus, in this log-log plot an object with zero internal extinction would be located at (0,0), which is indicated by dashed lines. 
The colored solid lines in the figures show the reddening vectors for a range of 1.4 magnitudes in $A_V$ for four popular extinction laws as indicated in the captions.

%$T_e>12000$ as measured from the [OIII]4363 and [OIII]4959,5007 lines.

%\begin{figure*}
%\includegraphics[width=0.4\textwidth]{balmer3.pdf} 
% \vspace*{-3.0cm}
%\caption{\small Same plot as Figure~\ref{two} .}
%\label{tres}
%\end{figure*}
 
While for the full sample of objects (gray points) the best fitting
extinction law appears to be unconstrained by our measurements, when
we restrict that sample to objects with errors $<0.015$ (blue
crosses), the best fit appears to disfavor the law of \cite{cal2000}
which produces much less reddening per unit absorption (it has
$R_V=4.88$).  Therefore, we have corrected our spectroscopic
observations using the Balmer Decrement with a Gordon extinction law
as

\begin{equation}
  \frac{\log {\cal F}(\lambda)_0}{\log {\cal F}({\rm H\beta})_0} =
\frac{\log {\cal F}(\lambda)_{obs}}{\log {\cal F}({\rm H\beta})_{obs}} -
  {\rm C(H\beta)}\times  f_\lambda 
\end{equation}
\noindent
where ${f_\lambda}$ is derived from the extinction law.  The resulting logarithmic extinction correction factor C(H$\beta$) for our sample is shown in a histogram of Figure~\ref{sample}.

%to be between the standard Galactic reddening law, and the one by \cite{cal2000}. Notice, however, that the latter produces much less reddening per unit absorption (it has $R_V=4.88$).

\subsubsection{Oxygen abundances}

We have determined the physical conditions from the emission line spectra by using the direct method as described by the prescription of \cite{pagel1992} and \cite{izotov2006} since we were able to detect and measure the electron temperature sensitive emission line of [O III]$\lambda$ 4363 in virtually all of our spectra (see Figure~\ref{sample}, right panel).  Electron densities were estimated by the ratio of the [SII] lines.  This[SII]$\lambda$6717/[SII]$\lambda$6731 ratio has a strong peak at 1.3 for our sample which implies that all HII regions are in the low density regime. Thus, we adopt the reasonable assumption of a constant n$_e$ of 100 cm$^{-3}$.

The ionic and total abundances determined by both prescription agree very well, with a small offset of less than 0.1 towards higher abundance for Izotov prescription which we adopt here for having more recently updated atomic data.   The prescription is also better suited for sub-solar abundances. While the low abundance tail in the distribution is expected to be real and accurate, the few objects with super-solar abundance have  larger uncertainties, due to the lower S/N ratio of the [OIII]$\lambda$4363 line in these cases.

\section{Modelling with CIGALE}\label{cigale}

\begin{table}
	\caption{The CIGALE Module parameters and their ranges for SED modelling.}
	\smallskip
    \centering
	\begin{tabular}{l c }
	\hline
    \noalign{\smallskip}
	Parameter & Value \\ 
	\noalign{\smallskip}
\hline
 \noalign{\smallskip}
\hline
 \noalign{\smallskip}
 SFH         			&  3 bursts 		        \\
 Age$_{young}$ [Myr]&  1, 2, 3, 4, 5, 6, 7, 8, 9, 10 \\
 Duration$_{young}$ [Myr]& 1, 2, 3, 4, 5 \\
 Age$_{int}$ [Myr]& 100, 500, 1000 \\
 Duration$_{int}$ [Myr]& 10,100 \\
 Age$_{old}$ [Myr]& 10000 \\
Duration$_{old}$ [Myr]& 100, 500 \\				
\hline
	Stellar Population Models & BC03 (1)	\\
	IMF & Chabrier\\
	Metallicity & 0.008\\
        \hline
        Nebular emission & \\
        Ionization Parameter & logU = -2.0 \\
        LyC escape & 0.0 \\
        LyC absorbed & 0.0 \\
\hline        
Dust attenuation & Calzleit (2) \\
E(B-V)$_{young}$	&  0., 0.05, 0.1, 0.15, 0.2, 0.3, 0.5	\\
E(B-V)[Young/Old] & 0.44, 1 \\
\hline		
Dust template & dl2014 (3)\\
Mass fraction of PAH & 0.47, 1.12 \\
Minimum radiation field & 0.1 \\
IR power-law slope ($\alpha$\tablefootmark{b})	& 	2.0  \\  
\hline
     AGN template & NONE \\
     \hline
     Radio & NONE \\ \hline
     Number of models run per & \\
     redshift bin ($\Delta$(z) = 0.01):  & 621600\\
	\noalign{\smallskip}
\hline
        \end{tabular}
    \label{cig:mod}

    \tablebib{
(1)~\citet{bc03}; (2) \citet{cal2000} \& \citet{leit2002}; (3) updated models of \citet{dl2007}.
}
    \end{table}

\begin{table}
  \centering
    \caption{Models: Input Data Set and Extinction Choices.}
    \begin{tabular}{lll} \hline
      Model & Dust emission & Extinction \\ \hline
 0 & No     &  free fit                    \\  
 1 & dl2014 &  free fit                      \\
 2 & dl2014 &  fixed H$\alpha$/H$\beta$ (Gordon $C=1.00$)\\
 3 & dl2014 &  fixed H$\alpha$/H$\beta$ (Calzetti $C=0.44$)\\
 4 & No     &  fixed H$\alpha$/H$\beta$ (Gordon $C=1.00$)\\
 5 & No     &  fixed H$\alpha$/H$\beta$ (Calzetti $C=0.44$)\\ \hline

    \end{tabular}
    \label{models}
\end{table}

\begin{figure*}
  \centering
  \includegraphics[width=0.49\textwidth]{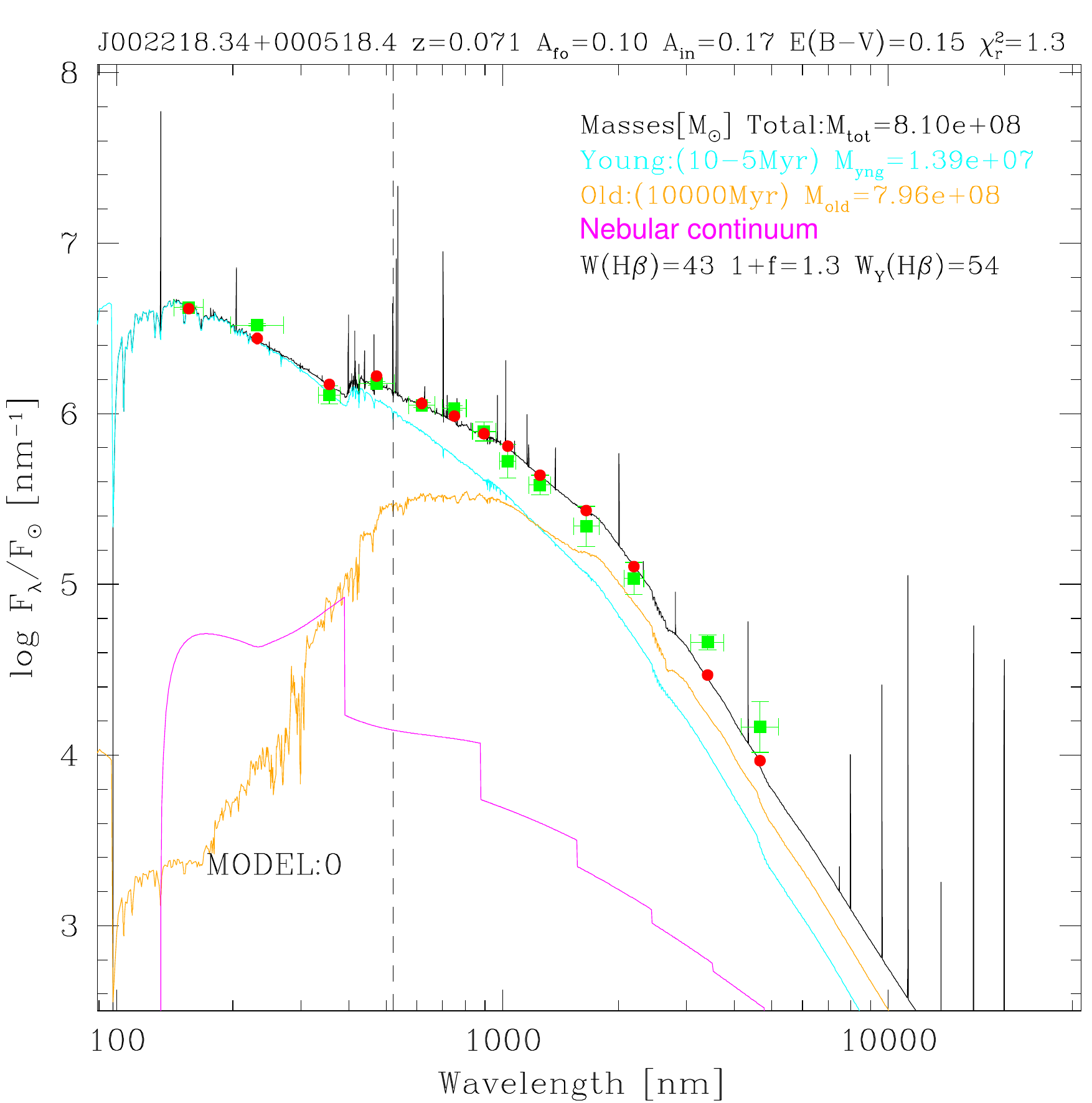}
  \includegraphics[width=0.49\textwidth]{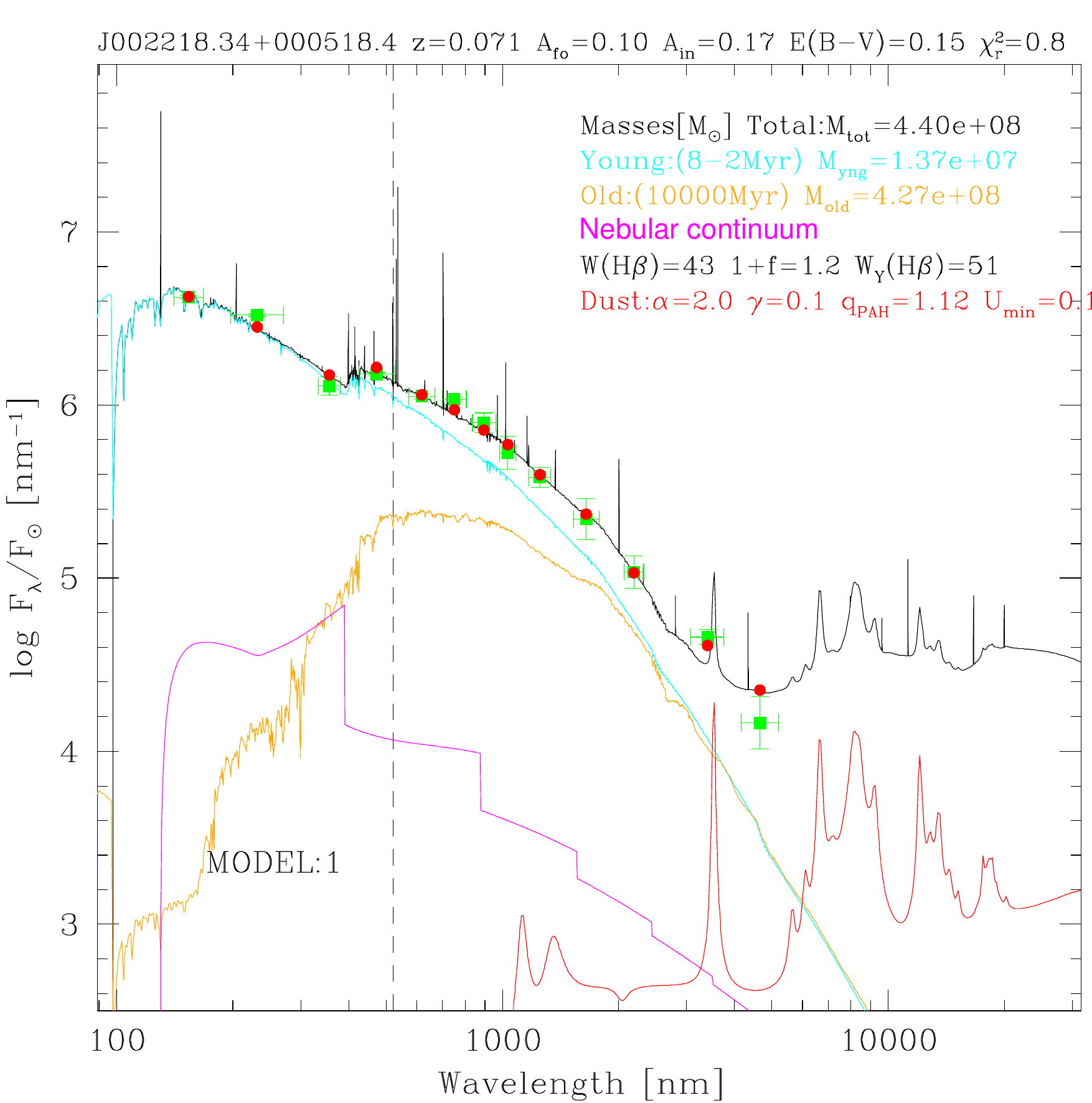} 
% \vspace*{-3.0cm}
\caption{\small Comparing free fit without dust emission (left panel, MODEL 0) vs. with dust emission (right panel, MODEL 1). The Green points are the observed photometry from UV to mid-IR. Only the first two WISE data points are shown and used in the fits. The solid lines are the modeled components: young stellar population (cyan), intermediate+old stellar population (orange), nebular continuum (magenta), and for MODEL 1 (right panel) the dust emission (red). The red points represent the model fit results for each photometric band. The information in the inset are the respective ages and derived masses of the stellar populations. The inset also shows the observed equivalent width of H$\beta$ (WH$\beta$)  and the corrected equivalent width of H$\beta$ for the young stellar component only (W$_y$(H$\beta$), see text.)}
\label{mol0mol1}
\end{figure*}

CIGALE\footnote{CIGALE software and documentation are available at: http://cigale.lam.br/}(Code Investigating GALaxy Emission) is a package for SED modelling as well as for SED fitting. The code has been developed by Denis Burgarella and M\'ed\'eric Boquien at Laboratoire d’Astrophysique de Marseille.  Some applications and descriptions of CIGALE to modelling galaxy properties are found in \cite{noll2009,boquien2014,boquien2016,ciesla2015,ciesla2016,vika2017}.
%\sout{A comprehensive description of CIGALE is also given in Noll et al. 2009.}
CIGALE has a modular structure which allows great flexibility in {\it modelling } the various physical components and their possible parameters that contribute to produce the theoretical SED of galaxies.  These components and the set of parameters used in our study (a subset of all possibilities) are given in Table~\ref{cig:mod}. Once the predicted theoretical SED are modelled, CIGALE is used for {\it fitting} the modelled SED to the observed SED.  Best fit results, probed by the output $\chi^2$ , can be evaluated to provide the possible and most probable set of energy sources and their respective parameters that best represent the observed SED.

CIGALE allows for a number of star formation histories (SFH) such as exponentially declined, delayed or periodic (see CIGALE documentation$^3$). Our choice of  the SFH consists of 3 episodes of star formation: a young ionizing population ($< 10$ Myr), an intermediate age population of (100-500 Myr), and an old stellar population (10 Gyr).

%Optical \citep{Telles1997a,Telles1997b} and near-IR imaging \citep[and references therein]{Lagos2011} have also convincingly shown that starbursts, and also dwarf starbursts, possess underlying old populations. Simulations also show the episodic nature of starbursts, particularly at low masses \citep[see e.g][]{Pelupessy2004, Debsarma2016}, being 3 episodes the simplest choice in this scenario.

In CIGALE, this particular SFH module was not implemented by default, but was
developed by M. Bocquien to fit our purpose. Once we chose the SFH, we also made a choice of the evolutionary stellar population synthesis to produce our Simple Stellar Population (SSP). We used the models of \cite{bc03} for a Chabrier Initial Mass Function (IMF) and at a fixed metallicity of Z=0.008.
The choice of metallicity is justified by the fact that we have derived, from the optical spectra, their low metal content with a mean distribution of 12+logO/H= 8.2 as shown in our  Figure~\ref{sample}.  These are typical values for HII galaxies where the electron temperature sensitive line [OIII]$\lambda$4363 is detected \citep{Kehrig2004,Ter1991}.  The oxygen abundances are expected to be low for our sample of high excitation HII galaxies, since they were selected to fall in the upper left locus of the star forming sequence in the BPT diagram, which is also a metallicity sequence \citep[see e.g.][]{curti2017}.

Nebular emission from the ultraviolet to the near-infrared was computed by the module Nebular that includes both emission lines and the nebular continuum. Here, for the computation of the nebular emission, we considered that the escape fraction and the absorption by dust of  Lyman Continuum were both zero. We may evaluate {\it a posterior} whether these choices were appropriate.

In all CIGALE runs we have not included a possible AGN
emission component.  Our targets are selected to be star forming
galaxies in the BPT diagram.  It is true that AGN at low mass and low
metallicity may exist and may fall in the star forming region of the
BPT \citep{stasinka2006}, but their frequency in our sample is
expected to be low, if any at all.

%One difficulty of deriving the attenuation by dust from a stellar population analysis is that we do not know a priori the details of history of SF, IMF, SFR, metalicities of the intrinsic spectra of the galaxies, as is the case for individual stars for the derivation of the extinction curves.  Therefore, we have to approach the problem in a statistical way with a large sample of galaxies, particularly the ones with the supposedly the simplest star formation histories, as in the case of local HII galaxies.  

Our choice of models differ simply on the way we considered the dust attenuation and the inclusion (or not) of dust emission in order to fit the whole spectral range from FUV to W2 band. Table~\ref{models} shows these six runs of CIGALE. Column 1 is the model identification, column 2 indicates whether or not a dust emission component was included in the runs. These are models 1, 2 and 3, marked ``dl2014'' indicate that the model of \cite{dl2014} was used.   The inclusion of dust emission will not affect the results significantly since dust emission becomes important only with $\lambda > 15 \mu$m.
%These are models 1, 2 and 3, marked ``dl2014'' in the 2nd column of Table~\ref{models}.  

The comparison between model 0 (free attenuation and no dust emission)
with model 1 (free attenuation with dust emission) allows us to
evaluate how much the inclusion of the dust emission interferes on the
full fit.
Figure~\ref{mol0mol1} shows a typical example of this case.  On the left panel the best fit for model 0 and on the right panel the best fit for model 1.
We can note that model 1 has a smaller $\chi^2$ that shows it is a better fit to the full wavelength range including W1 and W2 Wise bands.  In general, the routine tries to compensate the absence of the dust emission component in model 0 by over-estimating the mass of the old stellar population.  In this particular case the PAH 3.3$\mu$m emission also contributes to the flux in the W1 band. We conclude, then, that this dust emission component is necessary for any good fit in mid-infrared band, and contributes to a better fit in the near-infrared UKIDSS bands, as shown in this figure.

By applying analogous comparisons with the other models for which we had a choice of inclusion or not of a dust emission component, but with all other parameters being the same, such as in the case of model 2 vs. model 4 and model 3 vs. model 5 (see Table~\ref{models}) we reach similar conclusion, namely that models with dust emission are a better fit to the near-infrared data.  So from this analysis we have, then, favored the models where dust emission is included.  Therefore, we will discard models 0, 4 and 5 for the analysis that follows.

Now, we have to compare models where the attenuation corrections were left as a free parameter (model 1) to the models where the attenuation corrections were applied prior to the SED fitting procedure (model  2 vs. model 3,  see  3rd column of Table~\ref{models}).  A choice of Gordon extinction law was used in all SED fitting runs. 
As mentioned above, all data have previously been corrected for foreground extinction due to our Galaxy.  CIGALE will output the best fitted extinction parameter $C$ ($C=\frac{E(B-V)_{star}}{E(B-V)_{gas}}$) which represents the relation between the attenuation in the stellar continuum to the extinction in the nebular emission, to be either a Calzetti differential extinction ($C=0.44$) or an equal extinction in the two emission components ($C=1$).  CIGALE uses this parameter as the ratio the attenuation in the stellar continuum of the {\it old} stellar population to the stellar continuum of the {\it young} stellar population, not to the nebular emission. Massive young stars are embedded in the ionized HII regions, but the distribution of the young population, the older stellar population and dust may be related. However, it is not clear how these relations behave.  We prefer to constrain our model to the information from the nebula emission, fixing the total amount of extinction provided by the spectral information. In addition, a comparison of the resulting $\chi^2$ for the different models reveals marginal differences.  Model 2 has a distribution of $\chi^2$  marginally better than model 1. So for these two reasons we have chosen to keep model 2 and discard model 1.

For the other remaining models (2, 3) we pre-corrected the internal extinction  using the balmer decrement (H$\alpha$/H$\beta$) with the assumption of a Gordon extinction law with $C=1$ (models 2) or Calzetti extinction law with $C=0.44$ (model 3), as mentioned in Section~\ref{data}.
For these models we expect that CIGALE will output simply a residual extinction, since data were previously corrected for total extinction.  The smaller this residual, then, will indicate a better assumption of the attenuation law and of $C$.

\begin{figure}
  \centering
  \includegraphics[width=0.45\textwidth]{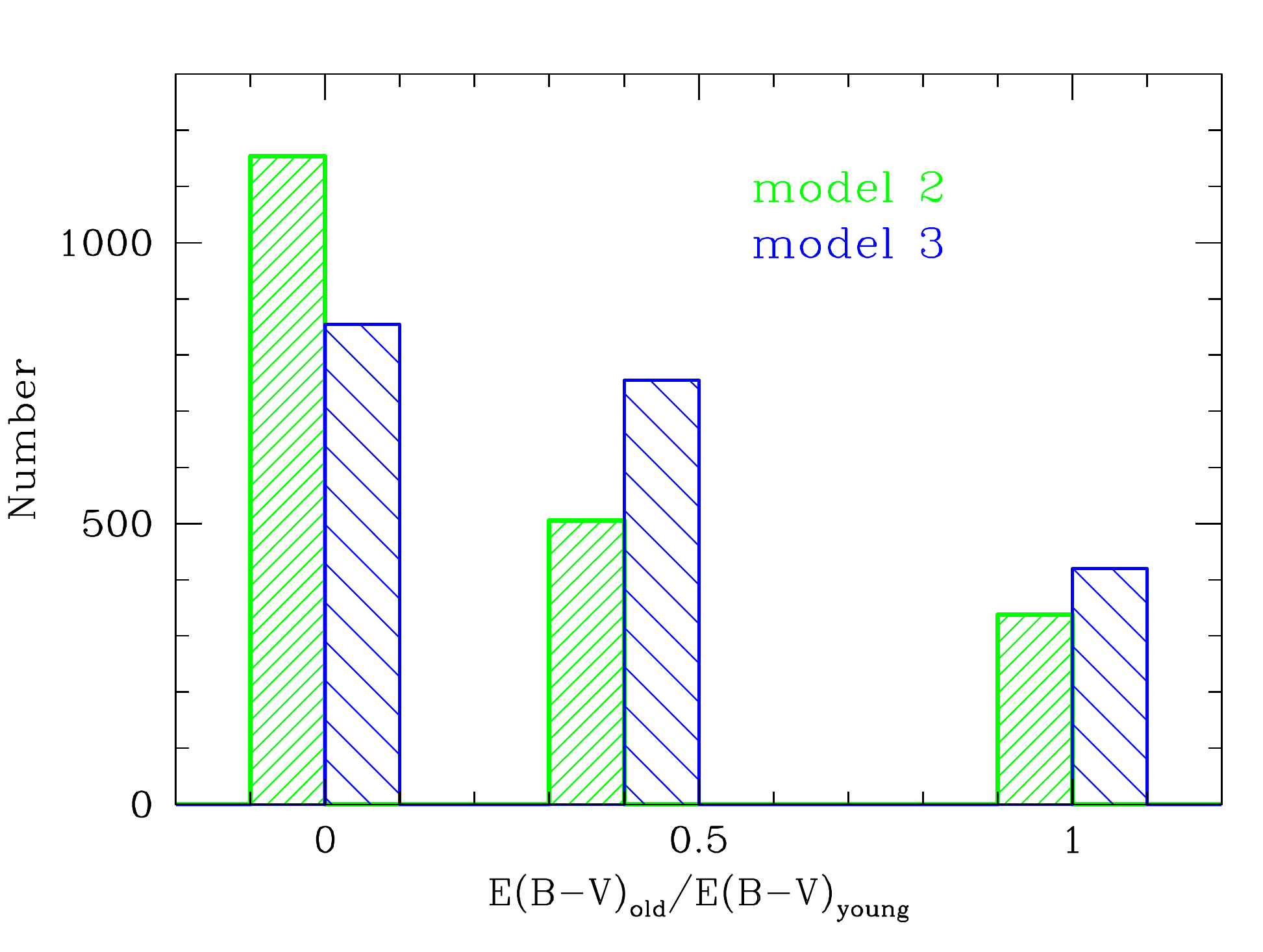}
\caption{\small Comparison of model 2 and 3 in relation to  the residual of best fitted  extinction parameter $C$ ($C=\frac{E(B-V)_{star}}{E(B-V)_{gas}}$). Model 2 has been pre-corrected for a Gordon C=1.00, and model 3 has been pre-corrected for a Calzetti C=0.44. Ideally, the assumption of a pre-correction would be perfect if this procedure resulted in zero residuals, which is not the case.  But simply model 2 (green) is better because it has relatively more zeros than model 3 (blue).}
\label{C_comp}
\end{figure}

Figure~\ref{C_comp} shows a comparison of the output results of CIGALE for $C$ for models 2 (green histogram) and 3 (blue histogram).  This represents the residuals of attenuation that CIGALE still manages to fit to find the best result. Model 2 has more zeros than model 3 and also more zeros than other residuals (either 0.44 or 1.0). This is not the case for model 3.
This is an indication the best prior extinction correction was performed using the assumption of model 2, using a Gordon attenuation law with no differential attenuation between gas and stars ($C=1$).

In summary, we have chosen model 2  as our best general model for the SED fitting procedure.  Our results will then be derived from the SED fitting using model 2 only (see Table~\ref{models}). In addition, only best fits with  $\chi^2 < 3$ will be considered for further analysis.

\section{Results and Discussion}\label{results}

\begin{figure}
  \centering
  \includegraphics[width=0.49\columnwidth]{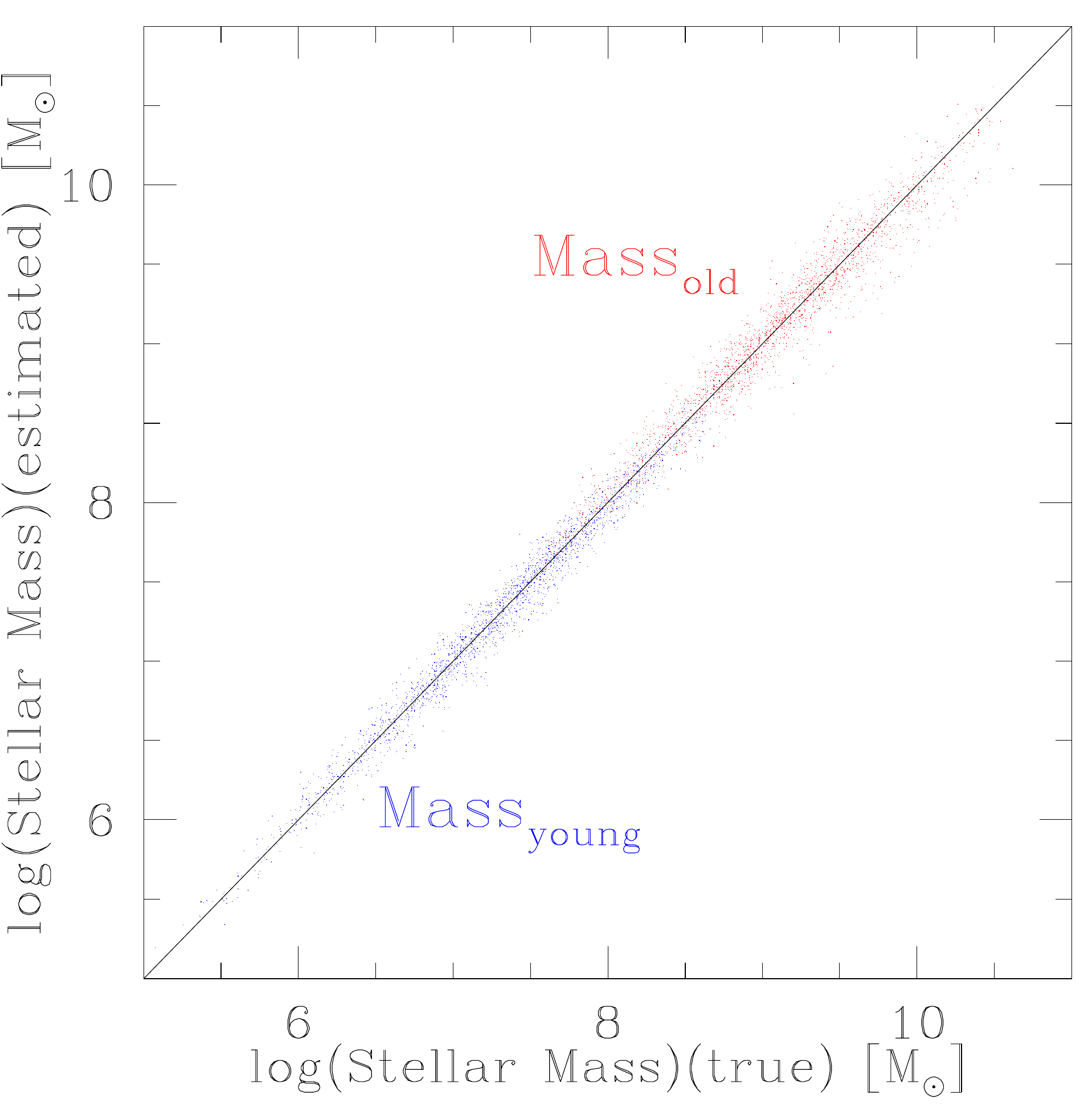}
      \includegraphics[width=0.49\columnwidth]{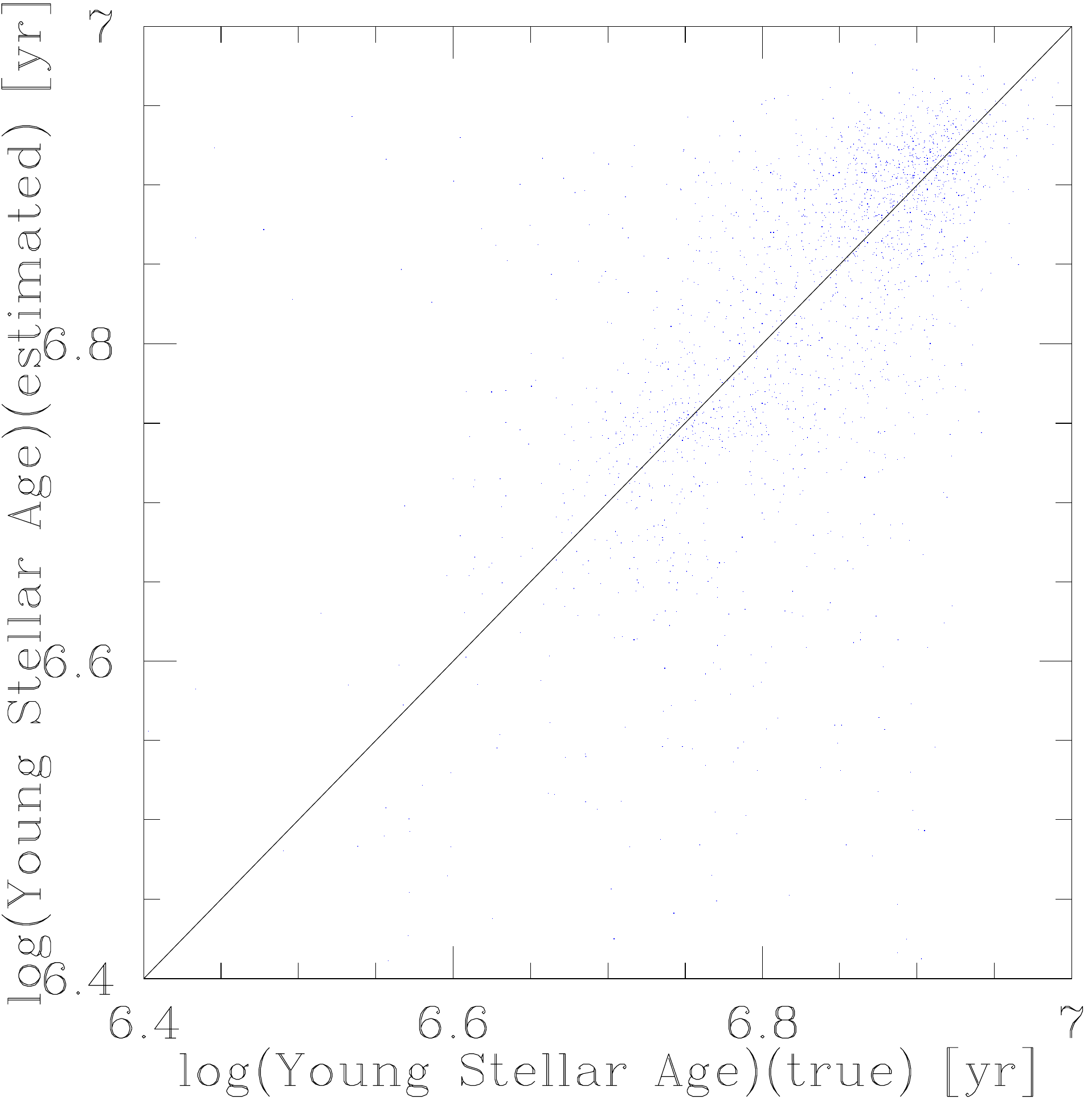}

\caption{\small Comparison between the input true parameters with the output estimated parameters from the mock catalogue created by CIGALE. Only the two main parameters of immediate interest are shown (mass and age). (left) Stellar Masses: Blue points are the masses of the young stellar component. Red points are the masses of the old stellar component (in our case Mass$_{old}$ + Mass$_{intermediate}$). (right) Young Stellar Age. The solid line  the 1:1 relation in both panels.}
\label{mock}
\end{figure}

As described in the previous Section,  our results are based on CIGALE model \#2 (see Table~\ref{models}), which includes dust emission for a better fit to the near-IR and WISE bands.  The photometry has been corrected for foreground (with a Galactic extinction law) and internal extinction (with a Gordon extinction law and E(B-V)$_{gas}$ = E(B-V)$_{star}$). CIGALE best-fit output parameters are the masses and ages (of the oldest stars) of the stellar components, the residual extinctions, the H$\beta$ equivalent width correction factor ($f_r$), and the best-fit $\chi^2$.
The realibitily of the output parameters by CIGALE has been tested in previous studies with a method by \cite{giovannoli2011} \citep[see also][]{buat2011} which consists of the creation of a mock catalogue of fluxes for each galaxy derived for a set of output parameters. After the addition of random noise CIGALE is run a second time and the new results are compared with the input ones.   \cite{vika2017} also applied this test for their analysis of Spitzer/IRS galaxies and showed that stellar masses, SFR, and luminosities are rather well constrained, whereas the age of the oldest stars is not. We have applied the same test here. Figure~\ref{mock} shows the comparison of our most relevant parameters for our present study: masses (left panel) and young  stellar age (right panel). We find that the stellar masses are very well constrained and reproduced.  In the right panel we show the comparison of the true vs. the estimated values of the age of the young stellar component only. The age of the intermediate stellar component shows a similar spread (not shown here). As a reminder, the age of the old stellar component is kept fixed at 10 Gyrs.  Note that the dynamical range in the plot for ages (right panel) is much smaller than the plot for the masses (left panel), so points look more spread.  However, in fact the reproducibility of young stellar ages also seem rather well constrained.

All the results from CIGALE, along with the derived spectroscopic  properties from the SDSS spectra, are available as FITS and ASCII tables at our website\footnote{http://staff.on.br/etelles/SED.html}. The most important output parameters measured by CIGALE, and used in our study are the stellar masses and ages of the young and old populations, and  the output best model SED with $\chi^2$, stellar, nebular and dust continua emission and respective attenuations. Plots (JPG images) of the resulting SED best-fit  for individual objects (as in Figure~\ref{mol0mol1}) are also made available, as well as the best fit model SED FITS tables.

\subsection{The Equivalent Width Correction (the $f_r$ factor)} \label{fr}

The observed equivalent width of H$\beta$, \eqw, has been shown to be an indication of the age of the burst \citep{dottori1981} for young stellar systems.

From a sample of HII galaxies, \cite{terl2004} showed that the observed distribution of \eqw\ cannot be reproduced if the evolution of the starburst is represented by a simple stellar population (SSP) predicted by evolutionary population synthesis models such as Starburst 99 \citep{sb99}.  The very high \eqw{$_{young}$} predicted for a very young stellar cluster is never observed, indicating that the observed \eqw{$_{obs}$} is actually a measure of the intensity of the H$\beta$ emission line (F(H$\beta$)) produced by the young burst averaged by the past history of star formation, including the continuum emission of the young massive star cluster (C$_{young}$) plus the continuum emission produced by the previous episodes of star formation (C$_{old}$). Hence,
\begin{equation}
{\rm EW(H}\beta)_{obs} = \frac{\rm F(H\beta)}{C_{young} + C_{old}}
\end{equation}
We have thus defined the fraction $f_r$ as the ratio  $f_r = \frac{C_{old}}{C_{young}}$ so,
%$C_{young} + C_{old} = C_{young} + C_{young} \times f_r = C_{young} (1 + f_r)$, then

\begin{equation}
{\rm EW(H}\beta)_{obs} = \frac{\rm EW(H\beta)_{young}}{1 + f_r}
  \label{eq:fr}
\end{equation}
%\begin{equation}
%  {\rm EW(H}\beta)_{young} = {\rm EW(H}\beta)_{obs} \times (1 + f_r)
%  \label{eq:fr}
%\end{equation}

\begin{figure}
  \centering
    \includegraphics[width=0.45\textwidth]{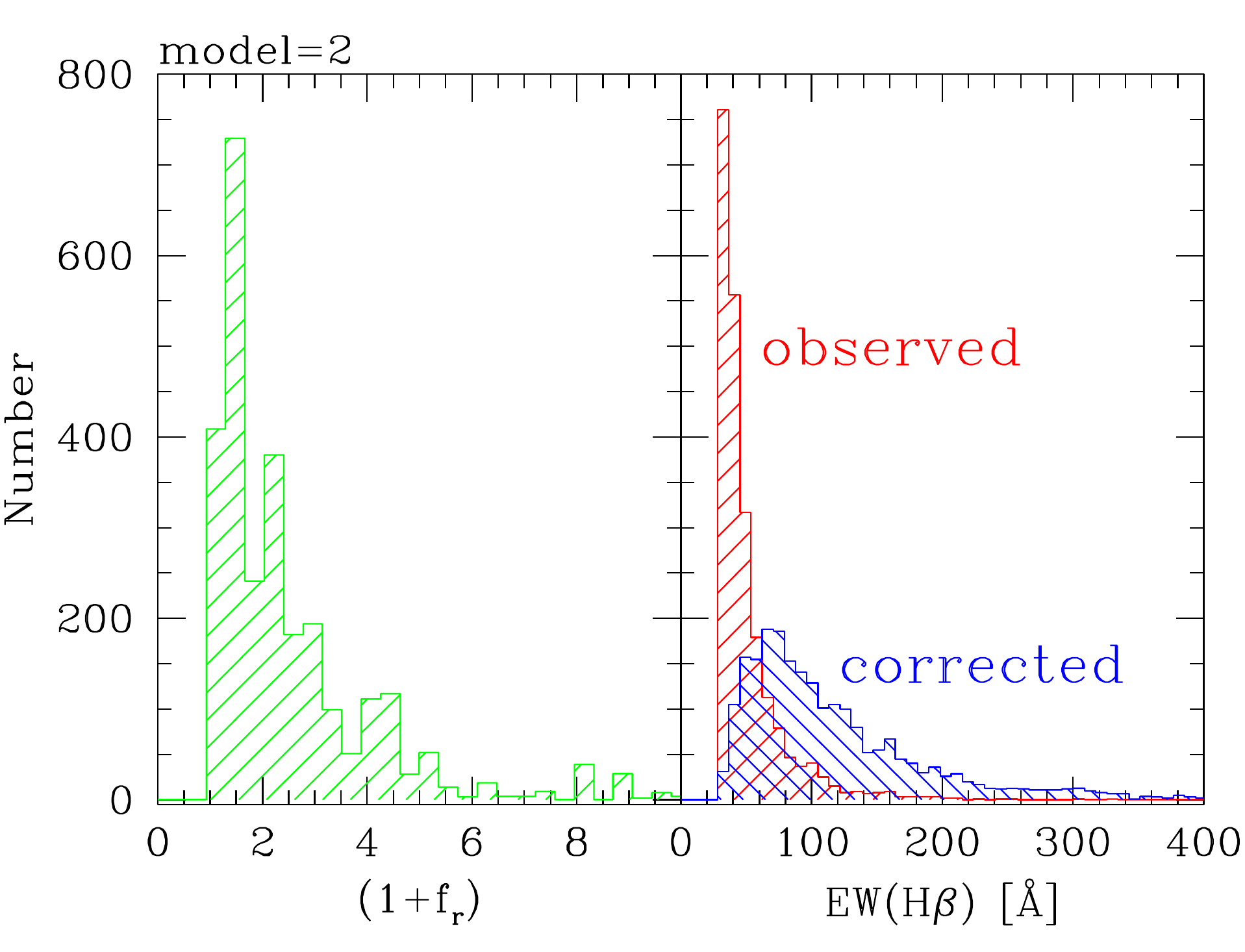}

% \vspace*{-3.0cm}
\caption{(left) ${\rm EW(H}\beta)$ correction factor $f_r$. (right) Histogram of the Distributions of equivalent widths of H$\beta$. The red histogram shows the observed values from SDSS spectra.  The blue histogram is the corrected EW(H$\beta$) for the contribution of an underlying old stellar population using the results from the SED fitting as described in the text.}
\label{ewb_hist}
\end{figure}

Figure~\ref{ewb_hist} (left panel) shows the derived  equivalent width correction factor $(1 +f_r)$ from equation~\ref{eq:fr}.  The resulting equivalent width distributions are given in Figure~\ref{ewb_hist} (right panel).  The corrected ${\rm EW(H}\beta)_{young}$ (blue histogram) is then the true equivalent width to be assigned to determine the burst ages, using equation~\ref{eq:fr} where $f_r$ is derived from the SED fitting.  The median value of the equivalent width correction factor  from equation~\ref{eq:fr} is $1 +f_r=2.0$.

\subsection{The relation between luminosity - and young stellar mass}

\begin{figure}[ht]
   \centering
\vspace*{-0.1cm}
\includegraphics[width=0.45\textwidth]{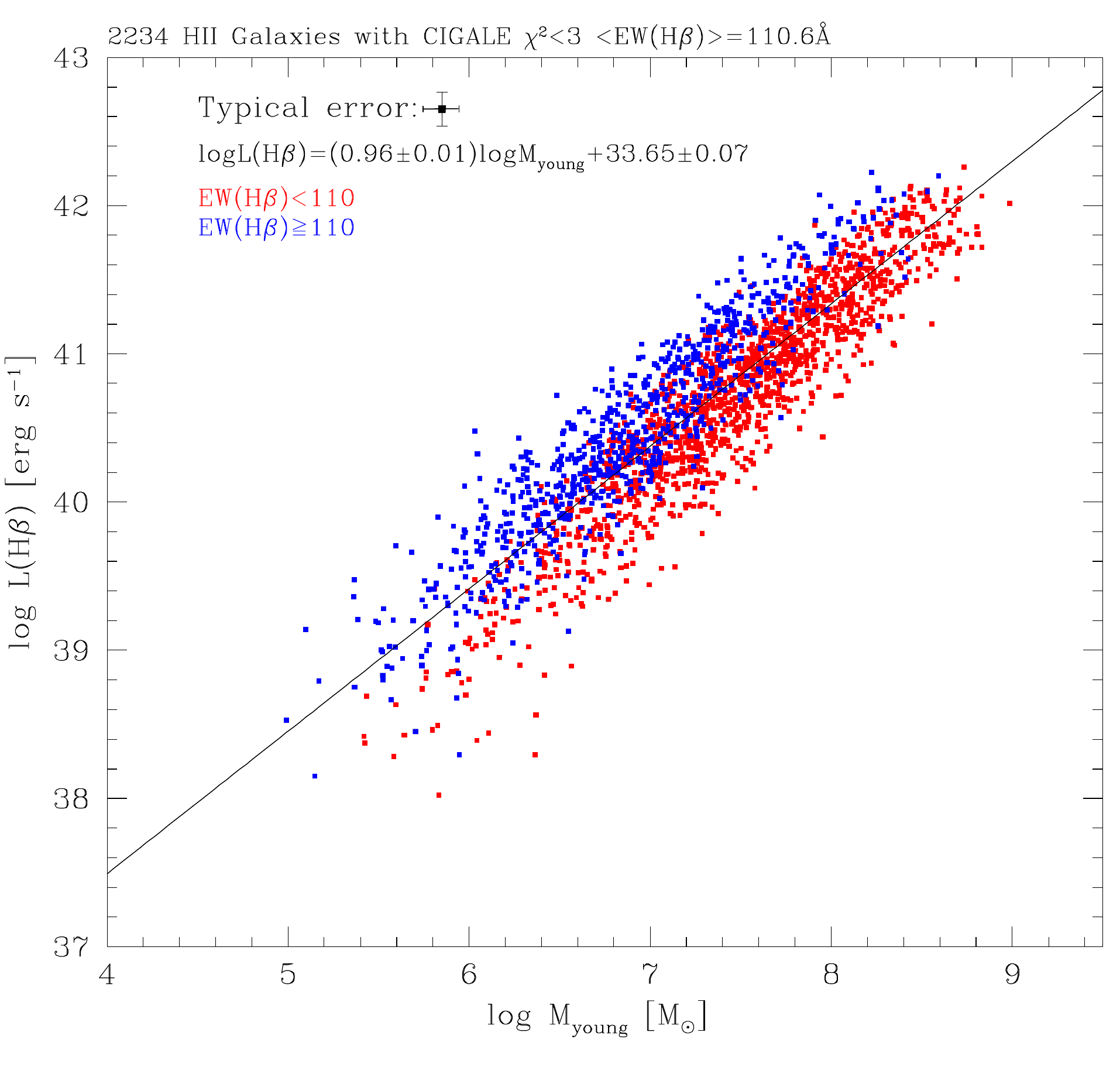}  
\vspace*{-0.1cm}
\caption{\small Relation between young stellar mass from CIGALE (\myou) and the observed \hbeta\ luminosity for our sample of 2234 HII Galaxies with
accurate SED fits ($\chi^2<3$). The sample was divided in two groups according to their equivalent widths, \eqw, corrected for contamination as described in Section~\ref{fr}.
Objects in red have \eqw\ lower than the average of the sample (110\AA) while objects in blue have values above the average. The line shows a standard least-squares fit of slope very close to unity as indicated in the legend. Typical error in \myou is $< 30\%$.}
\label{ML1}
\end{figure}

It is notoriously difficult to estimate the uncertainties of the parameters resulting from populations synthesis models.
Statistical errors in CIGALE are  estimated through  bayesian probability distribution functions and are given for each output parameter. Typical error for the masses in our work is 20\%-30\%, and for SFR errors are estimated to be 30\%-40\%. 
  A straightforward way of verifying our results is to compare the mass of young stars ($M_{young}$)  from CIGALE with the observed \hbeta\  luminosities, \lbeta. This comparison is presented in Figure~\ref{ML1} that shows an excellent correlation between these parameters. Notice, however, that since mass and luminosity depend on the square of the distance the slope of log-log plots such as this is expected to be close to unity even when the objects span a relatively small range of distances, so the interesting information is in the scatter and the zero point, but not necessarily in the slope.

The ionizing fluxes of single-age (simple) starbursts depend mostly on two parameters: the age and the mass of the ionizing stars,  and for these objects the equivalent width of the Balmer lines provides a robust age indicator (\citealt{sb99}, henceforth SB99). Thus, we expect the scatter in the relation between \myou\ and \lbeta\ to be correlated with \eqw. Figure~\ref{ML1} shows that this is indeed the case. Objects with \eqw\ lower than average lie predominantly below the fit line, while objects with values above average are above the line. Notice that the ridge separating these two groups is tilted relative to the least-squares fit.  

In the figure we used the equivalent widths corrected for contamination by old stars as described in Section~\ref{fr}, but the separation also occurs when the uncorrected \eqw\ are used, albeit with larger scatter and more overlap between the two groups (cf. Table~\ref{fifi} below).

\begin{figure}[ht]
  \centering
\includegraphics[width=0.45\textwidth]{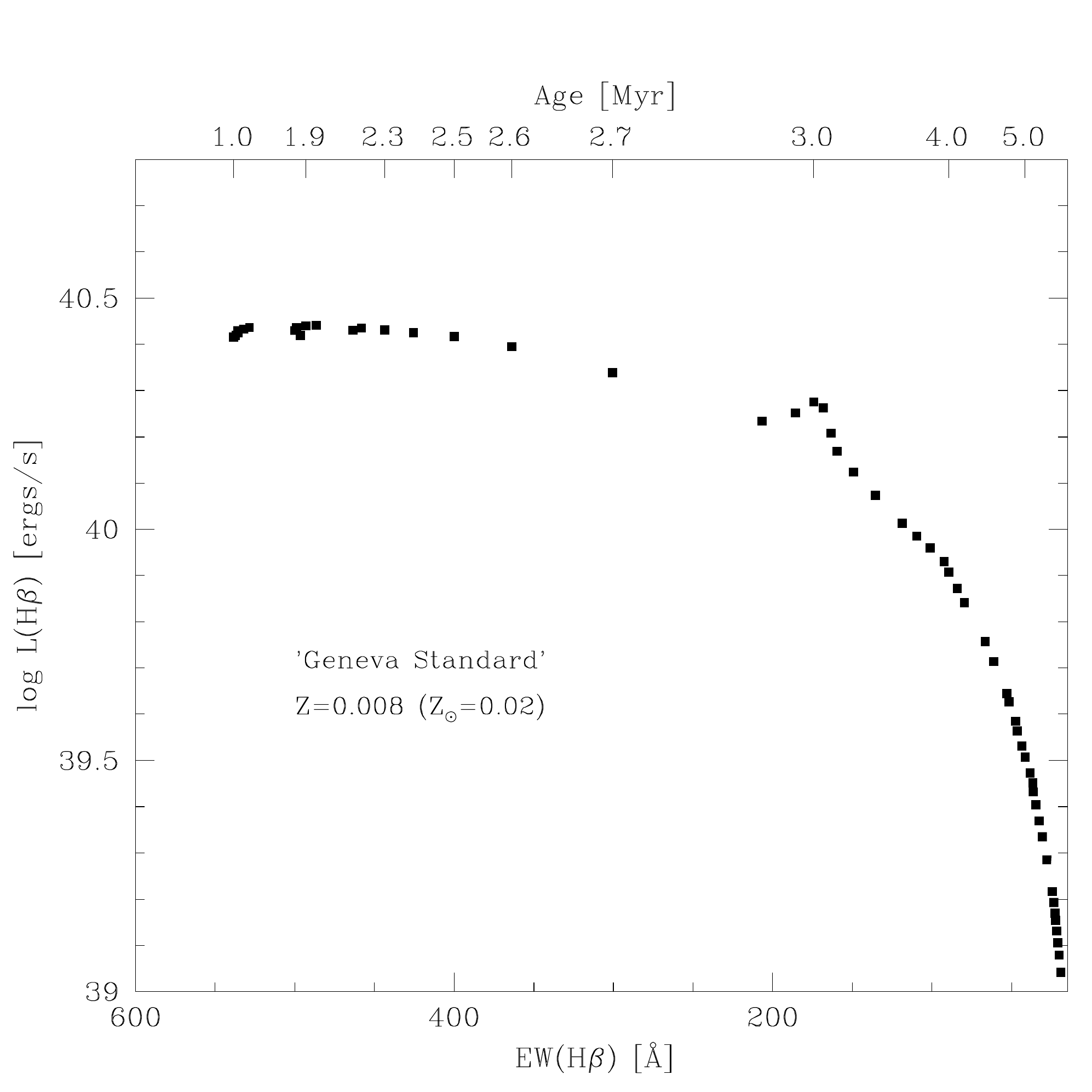}  
\vspace*{-0.1cm}
\caption{\small Relation between \lbeta\ and \eqw\ for a simple $10^6$\msun\ starburst of Kroupa IMF (from SB99). The age scale is shown at the top of the figure. }
\label{SB99}
\end{figure}

In \cite{mel2017} we showed how the SB99 models can be used to normalize the observed \hbeta\ luminosities to a fiducial age. Here we have used a dense grid of SB99 models for the standard Geneva isochrones with a metallicity of Z=0.008 and a Kroupa IMF shown in Figure~\ref{SB99}. This allows us to directly interpolate the models to the observed \eqw\ without recourse to fitting some analytic expression as done in \cite{mel2017} or \cite{Arenas2017}.

\begin{figure}[ht]
\centering

\includegraphics[width=0.45\textwidth]{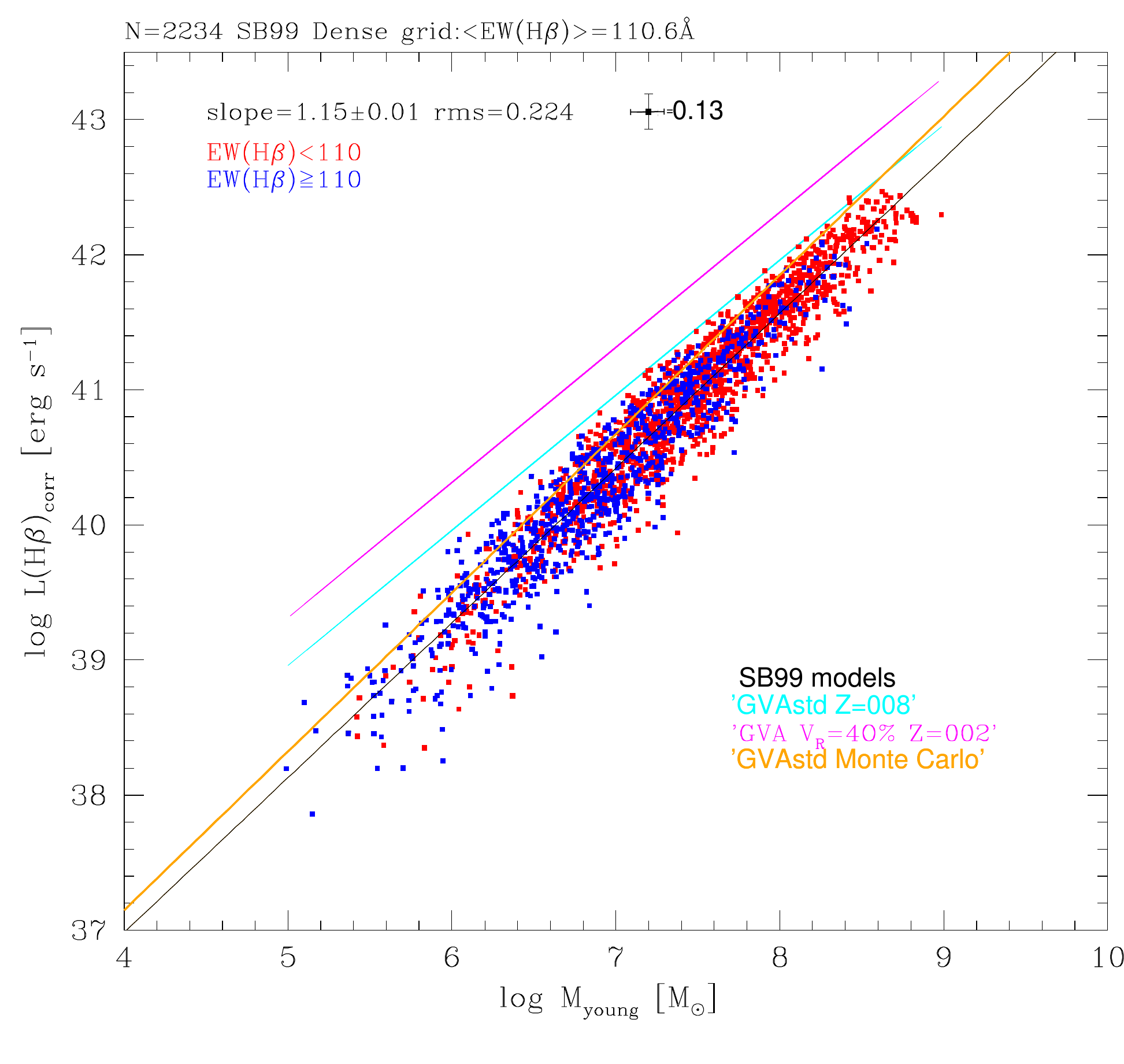}  
\vspace*{-0.1cm}
\caption{\small Relation between young stellar mass from CIGALE (\myou) and \hbeta\ luminosity at a fiducial age of 3.8Myr.  
As in Figure~\ref{ML1} the sample was split in two groups according to the corrected equivalent widths.
We show in red objects with less than the average of the sample (110\AA) and in blue objects with larger values. The coloured lines show the predictions of SB99 models for two different stellar models as indicated in the figure. The orange line corresponds to the Monte-Carlo sampling discussed in the text.}
\label{ML2}
\end{figure}

Thus, for each object in our sample we interpolate the SB99 models to retrieve the luminosity offset between the observed age and the fiducial age, for which we chose the mean age of the sample, and we scale the corrected luminosity to the actual mass (\myou) of the object. For the SB99 models that we are using here the mean equivalent width of our sample corrected for contamination by old stars $<EW(H\beta)>=110.4$\AA\,  corresponds to a mean age of 3.8Myr.  Figure~\ref{ML2} shows the relation between young stellar mass and \hbeta\ luminosity at a fiducial age of 3.8Myr.

The scatter is significantly reduced while the stratification of luminosities as a function of \eqw\ is gone. Interestingly, however, the figure shows a systematic trend of \eqw\ with mass: massive objects tend to have lower equivalent widths. The figure also shows that simple SB99 models predict significantly larger luminosities than observed, and that the relation using corrected luminosities is steeper than the uncorrected case (Figure~\ref{ML1}). We find, therefore, that single-age models provide the correct slopes ($dlogL/dEW$) of the evolutionary corrections, but not the correct zero points.

\begin{figure*}[ht]
 \centering
 \includegraphics[width=0.85\textwidth]{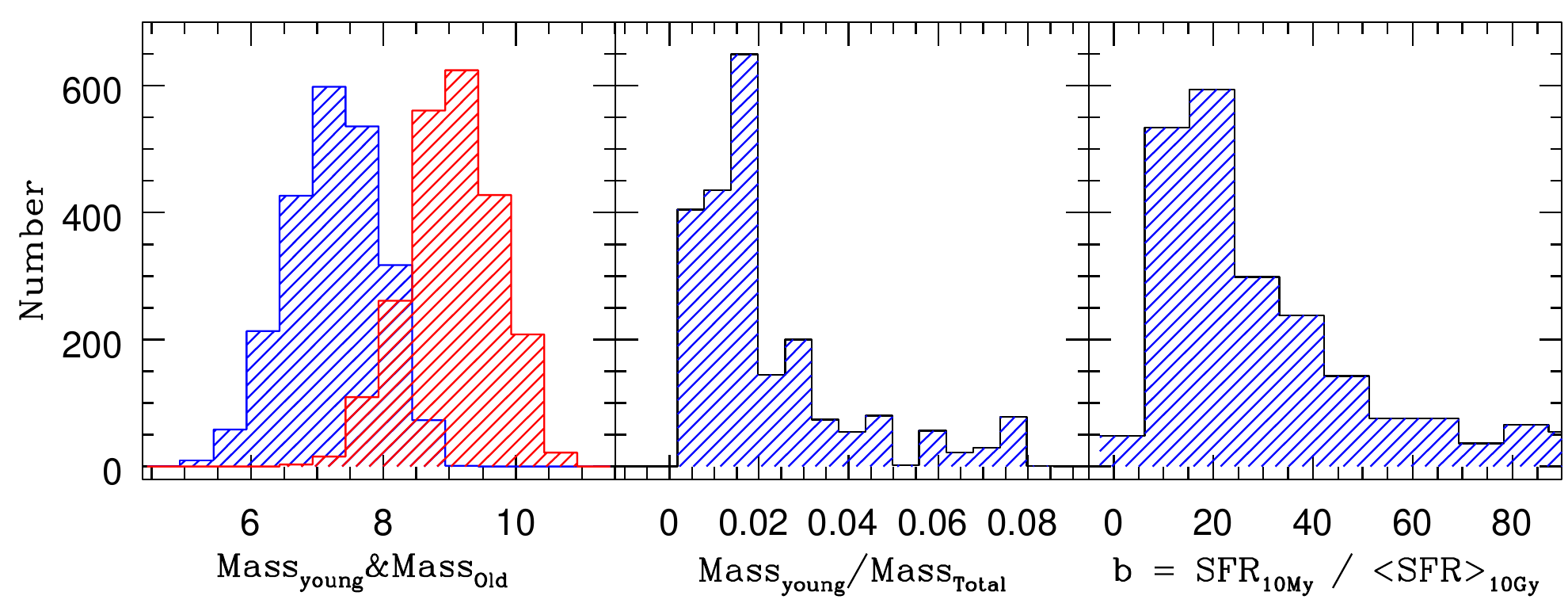}
%  \includegraphics[width=0.3\textwidth]{masY_masO_hist.pdf}
%  \includegraphics[width=0.3\textwidth]{mas_ratio_hist.pdf}  
%   \includegraphics[width=0.3\textwidth]{b_hist.pdf}
% \vspace*{-3.0cm}
\caption{(left) Young stellar mass (blue) \& old stellar mass (red) histograms. (center) Burst strength ($f_{burst} = \frac{M_{young}}{M_{old}}$).  (right) Birth rate parameter  \citep[$b = \frac{SFR}{<SFR>}$,][]{ken1983}, but in this case SFRs are from the SED fitting results $b_{burst} = \frac{M_{young}/Age_{young}}{M_{old}/Age_{old}}$. }
\label{b_hist}
\end{figure*}

Simple dynamical arguments indicate that very massive strictly coeval (simple) starburst clusters cannot exist. The characteristic time scales for the formation of such clusters would be too long compared to the main-sequence life-times of the ionizing stars. So it is reasonable to assume that the stars in massive starbursts span a range of ages that is significant relative to the ages of the ionizing stars. In fact, observations of nearby HII Galaxies show that these objects tend to be clumpy and have multiple emission-line profiles \citep{Lagos2011,mel2017}.

In order to test this multiplicity effect we performed simple Monte Carlo experiments where we split the young component of each galaxy with $M_{young}>3\times10^6 M_{\odot}$ into a set of clumps of masses $M_{cl}$ randomly sampled from a power-law mass distribution of slope $\alpha=-2$ in the range $3\times10^5<M_{cl}/M_{\odot}<3\times10^6$. We then assign to each clump a random age sampled from a Gaussian distribution centered at the mean age of the galaxy (from \eqw$_{young}$) with a dispersion that is a function of total young stellar mass: $\sigma_{age}=3\times(M_{young}/15)^{0.2}$Myr, with $M_{young}$ in units of $10^6$\msun. This generates a 3D grid from which we can read the predicted luminosity for a given mass and equivalent width. 

The orange line in Figure~\ref{ML2} illustrates the predictions from our simple Monte-Carlo sampling. The prediction has the right slope, but is still offset by about 0.3dex relative to the observations. It may be possible that more refined models could explain this offset, but a full sampling of parameter space is beyond the scope of this paper. For our immediate purposes, the important result is that simple SB99 models are  adequate for estimating the evolutionary corrections to the \hbeta\ luminosities of young starbursts.

In Table~\ref{fifi} we explore possible additional sources of systematic scatter in the \lhbeta-\myou\ relation. The evolutionary corrections parametrized by the raw equivalent widths is shown in the first line, and the corrected  evolutionary corrections for contamination by old stars in the second line. The rest of the table explores the scatter of the age-corrected mass-luminosity relation discussed above (Fig.~\ref{ML2}).  

We find a weak correlation with nebular excitation ([OIII]/[OII]), which is probably a residual from our evolutionary corrections. The decrease in scatter is deceiving because, as shown in Figure~\ref{ML2}, when the luminosities are corrected for evolution using the SB99 models the scatter is rms=0.22. Also, the 1965 galaxies with measured [OII]3727\AA\  tend to be those with the best S/N.  We do not see any residual correlation with metallicity or [OIII]/\hbeta. We conclude, therefore, that to a very good approximation, the \hbeta\ luminosities of HII Galaxies depend only on two parameters: the mass and the age of the starburst component. An immediate corollary of this conclusion is that the IMF of HII Galaxies is universal, at least for massive stars.

\begin{table}
  \begin{threeparttable}
  \tiny
  \centering
\caption{Multi-parametric Fits.}
\tabcolsep 1.5mm
\tiny
\begin{tabular}{lcccc}  
\hline\hline
		        					& \multicolumn{4}{c}{ $ \rm log L(H\beta) = c_0 +c_{1}log M_{young} + c_2X$ } \\ \hline
\ \ \ \ \ \ $X$ 					& $c_0$ 			& $c_1$ 			& $c_2$ 			& rms 	\\ \hline
$\rm EW(H\beta)_{obs}$	 		& $33.34\pm0.060$  & $0.963\pm0.008$	& $6.245\pm0.226$ 	& 0.263	\\ 
$\rm EW(H\beta)_{corr}$ 			& $32.45\pm0.069$	& $1.091\pm0.009$	& $2.325\pm0.075$ 	& 0.255	\\
$\rm log [OIII]/[OII]^1$\				& $32.67\pm0.060$	&$1.108\pm0.008$	& $0.029\pm0.010$	& 0.213	 \\ 
$\rm log [OIII]/H\beta^1$ 			& $32.40\pm0.007$	&$1.146\pm0.007$	& $0.005\pm0.035$ 	& 0.223	\\
$\rm 12+log(O/H)^1$				& $32.40\pm0.062$	&$1.147\pm0.008$	& $-0.001\pm0.003$	& 0.222	 \\ \hline
\label{fifi}
\end{tabular}
\begin{tablenotes}
  \tiny
\item   $^1$Using \lbeta\ corrected for evolution using the equivalent widths as in Fig.~\ref{ML2}.
  \item  To make the coefficients easier to read we scaled the equivalent widths by a factor of $10^3$.
%\hspace{10cm}  $^1$Using \lbeta\ corrected for evolution using the equivalent widths as in Fig.~\ref{ML2}.
\end{tablenotes}
\end{threeparttable}
\end{table}

	\subsubsection{The most massive starbursts}

We remarked above that Figure~\ref{ML2} shows a clear systematic decrease of \eqw\ with young stellar mass, in the sense that the most massive objects tend to have the lowest equivalent widths. The galaxies in our sample show a rather weak trend of metallicity with \myou\ for low mass objects while the most massive starbursts span the full range of metallicities, so we were puzzled by the fact that, even after correction for underlying older stellar populations, the most massive objects in our sample are still those with the lowest \eqw.
 
Visual inspection of the SDSS images revealed that most of these massive HII Galaxies show disturbed morphologies reminiscent of major mergers. Thus, the most luminous objects in our sample seem to be the low-mass equivalents of LIRGS and ULIRGS - the descendants of mergers of massive spiral galaxies.

It may be interesting to notice in this context that the relation between age dispersion and mass ($\sigma_{age}\propto M_{young}^{0.2}$) from our Monte-Carlo experiments is flatter than what we expect from the Virial theorem and the \lsigma\ relation, $\sigma_{age}\propto M_{young}^{0.25-0.4}$. This may be an indication that mergers rather than monolithic collapse controls the age spread in the most massive objects. Rejuvenation of massive stars through binary interactions may also play a role, although clearly more elaborate models are required to address these issues properly.

\subsection{Stellar Masses and Star Formation}

\begin{figure}
  \centering
  \includegraphics[width=0.49\textwidth]{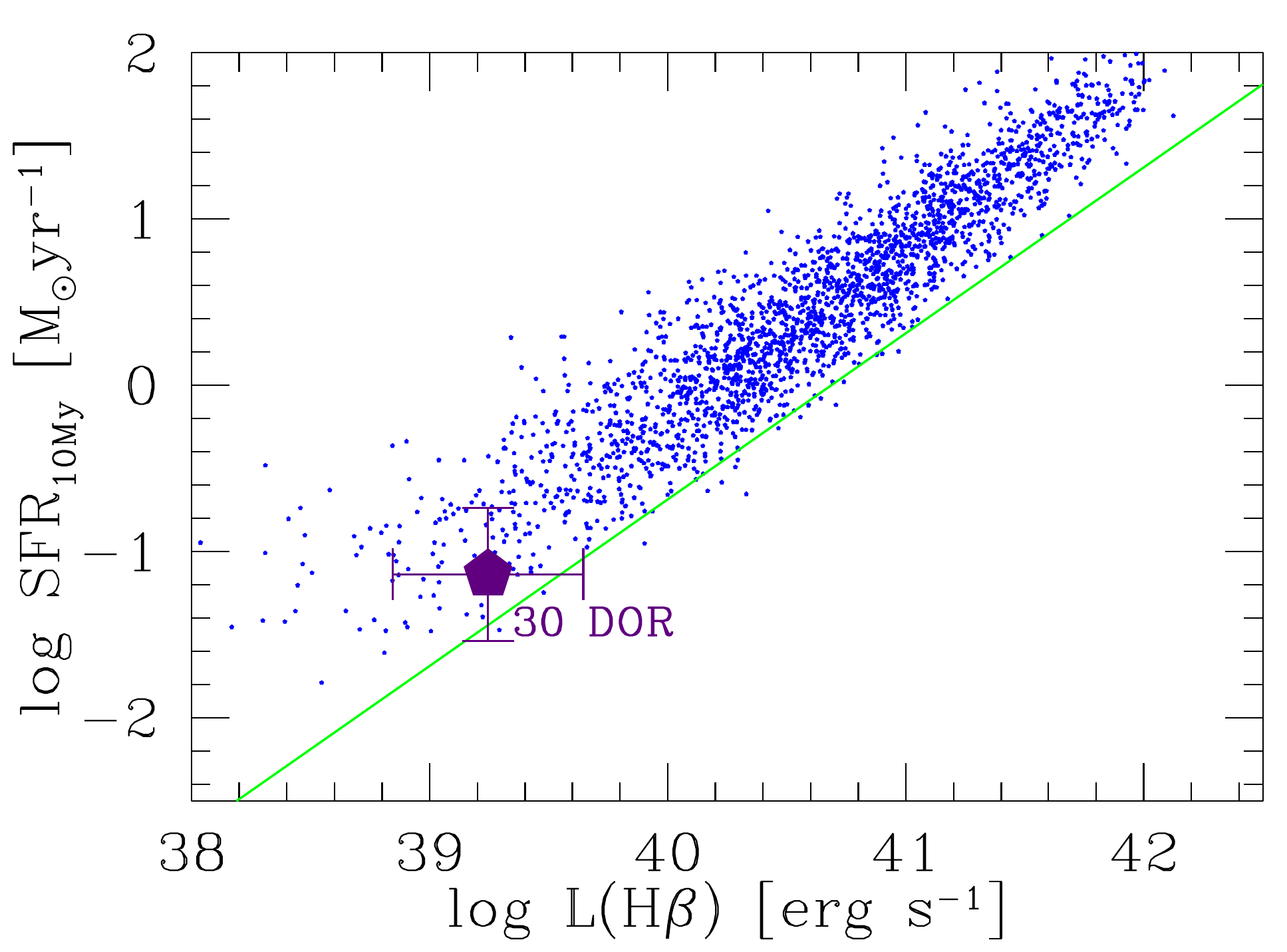}

% \vspace*{-3.0cm}
  \caption{SFR vs.  LH$\beta$ calibration. The blue points are for the current present-day star formation rate from SED result of SFR$_{10My} = M_{young} / age_{young}$ (<age>$_{young}$ = 6.8 My).
    The green line is the relation by \cite{ken98} for normal spiral galaxies.  The purple hexagon is the giant HII region 30 Dor in the LMC \citep{doran2013,crowther2017}.}
\label{sfr_lhb}
\end{figure}

Figure~\ref{b_hist} shows histograms of indicators of the importance of the current present-day star formation (SF) episode over the past  history of star formation from our SED fitting.  The left panel  shows the histograms of the derived masses for the young stellar component (M$_{young} <$ 10 Myr) to be compared with the old stellar component (M$_{old}$ =  M$_{int}$ + M$_{10Gy}$).  Notice that we are not making a distinction in our CIGALE models between mass produced in the intermediate-age episode (M$_{int}$) from the mass of first episode of star formation (M$_{10Gy}$) in our three burst model, so we refer to  these two components simply as M$_{old}$.

One can see that our galaxies have total masses of  less than 10$^{10}$\msun, and typically 10$^{9}$\msun, putting them in the low mass tail of other studies of overall properties of star forming galaxies.  This is of course due to our selection criteria as explained in Sec.~\ref{data}.

The middle panel shows the strength of the burst parameter defined as $f_{burst} = \frac{M_{young}}{M_{old}}$.  It is clear from this histogram that the present episode of SF has contributed less than a few per cent to the total stellar mass production over the lifetime of these galaxies -- typically less than 2\%. Analogously, the right panel shows the $b$ parameter \citep{ken1983}, here defined as ratio of the present-day SFR (SFR$_{10My}$) to the average past SFR (SFR$_{10Gy}$). The current star formation is producing stars at high rates, typically over $\sim$ 20 times the average past history.

\begin{figure*}
  \centering
  \includegraphics[width=0.49\textwidth]{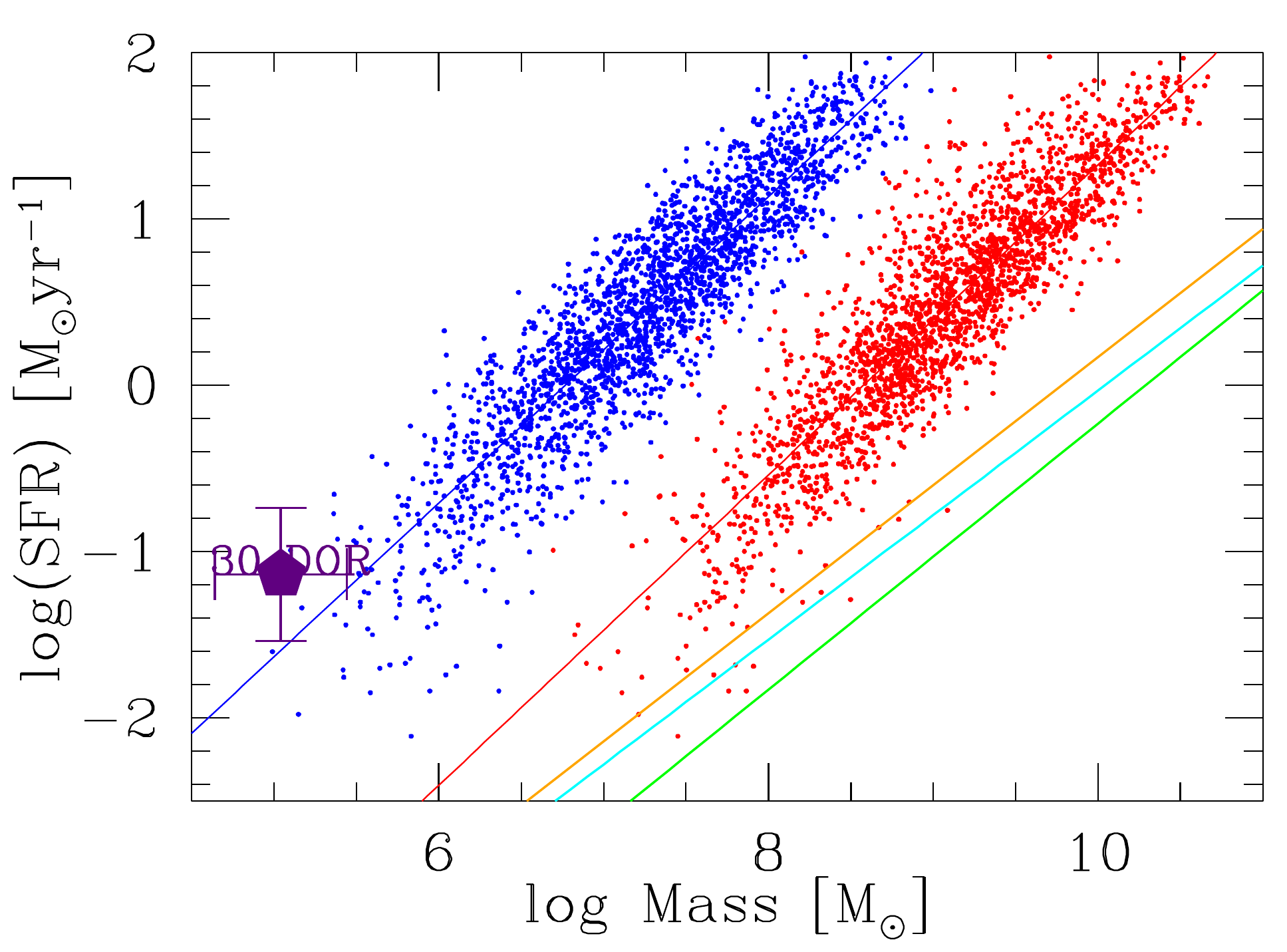}
  \includegraphics[width=0.49\textwidth]{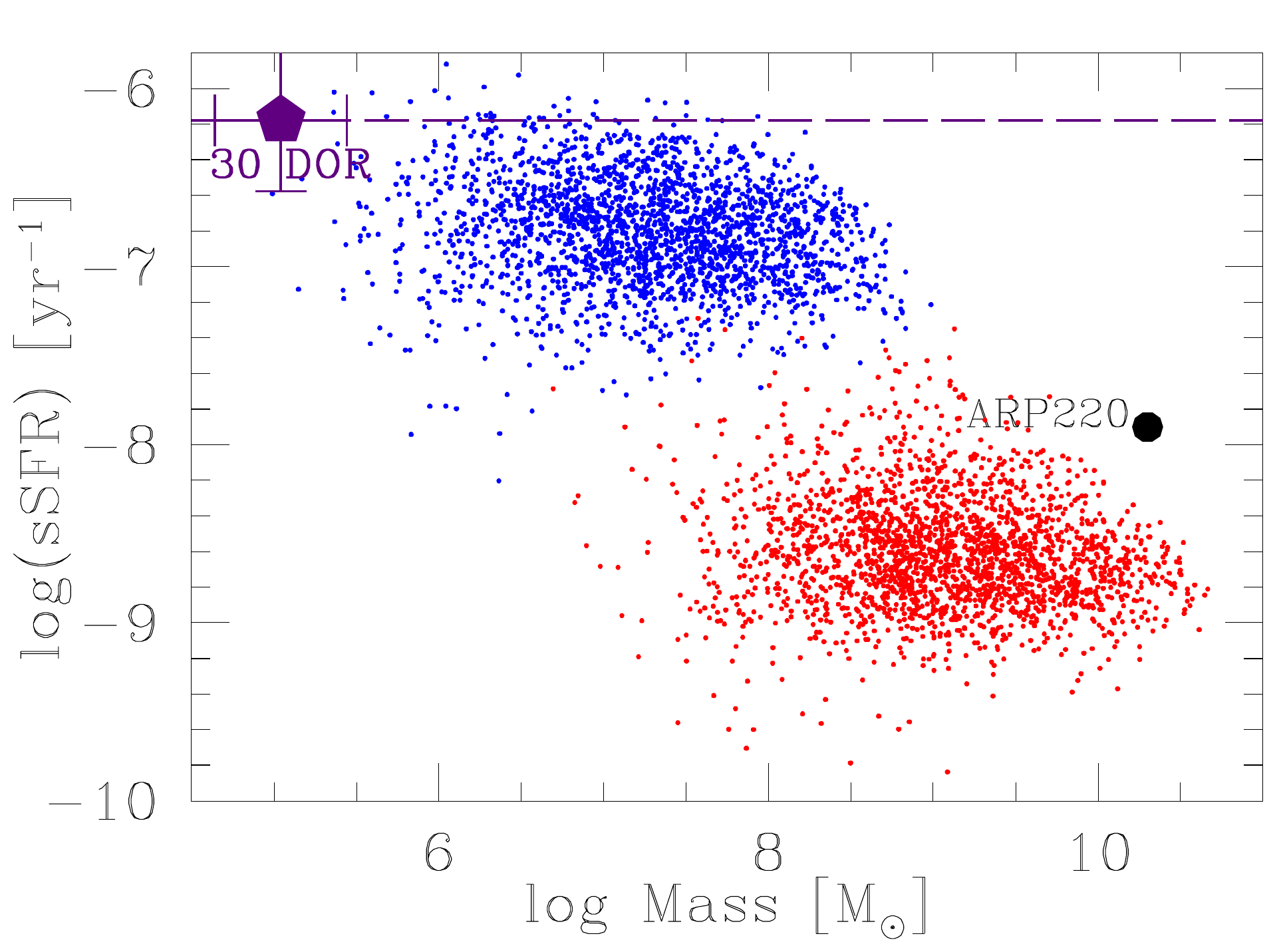}
% \vspace*{-3.0cm}
\caption{(left) The main sequence of HII galaxies: current SFR derived from the calibration between our SFR averaged over 10 Myrs from our SED fitting procedure vs. the LH$\beta$  from equation~\ref{sfr10my}, as a function of Mass$_{young}$ (blue points and line), and of Mass$_{total}$ (red points and line).  The green line is from \cite{brinch2004}, the cyan line comes from \cite{chang2015}, and the orange line comes from \cite{elbaz2007} (see text).  (right) Specific SFR as a function of Stellar Mass as in the left panel.  In both plots the positions of the genuine starburst 30 Dor in the LMC are given (magenta). The black point in the right panel represents the position of the ULIRG Arp 220 with SFR over $\sim$ 200\msun/yr. }
\label{ms_gals}
\end{figure*}

Our SED fitting procedure  allows us to isolate the mass of the young stellar component produced in the latest SF episode (M$_{young}$). Thus, the current SFR$_{10My}$  is  simply  M$_{young}$ / age$_{10My}$. We used the actual CIGALE best fitting young ages of the individual objects, although little differences would result had we used the mean age for our sample of <log(age)$_{10My}$> = 6.8, or a more conservative maximum age for the ionizing population of log(age)$_{10My} = 7$. For the SFR averaged over the whole SF history of the galaxy SFR$_{10Gy}$ we take M$_{old}/10^{10}$ yr.

Figure~\ref{sfr_lhb} shows our current SFR derived from our SED fitting procedure  $\log({\rm SFR}_{10My})$, plotted against our observed H$\beta$ luminosities corrected for extinction (see Sec.~\ref{data}). The resulting calibration forcing the slope to be exactly one is given by:
\begin{equation}
  \log({\rm SFR}) = -40.15\pm0.31 + \log {\rm L(H}\beta) ~(rms=0.25)
  \label{sfr10my}
\end{equation}

For comparison, the commonly used calibration of \cite{ken98}, shown as the green line in the figure, is  $\log({\rm SFR})= -40.65 + \log {\rm L(H}\beta)$. The difference in zero point is only partially due to a slightly different IMF, stellar input in the synthesis model, or aperture effects, and mostly to the fact that our calibration isolates the SFR of the starburst component alone. Thus,  the Kennicutt relation underestimates the present SFR  of starbursts typically by a factor of 3. 

The prototypical starburst 30~Dor in the LMC is seen in Figure~\ref{sfr_lhb} to fall exactly within the errors in the locus of HII galaxies confirming that the starbursts in HII galaxies are similar to the Giant HII Regions in local spiral and irregular galaxies.

The relation between SFR and stellar mass is generally known as the ``main sequence'' of star-forming galaxies. There are extensive discussions of this relation in the literature and its use as a probe of galaxy evolution as a function of mass, environment, and galaxy types \citep{perez2003,brinch2004,noeske2007,salim2007,daddi2007,elbaz2007,peng2010,chang2015,salim2016}, as well as sub-sets of extremely metal-poor galaxies \citep[and references therein]{filho2016} and samples of  BCDs \citep{sanchez2009,izo2014,janow2017}.  Various SFR indicators have been used but Kennicutt's  H$\alpha$ luminosity is the most common indicator for local star forming galaxies \citep{ken1983,brinch2004}.

Figure~\ref{ms_gals} (left panel) plots the relationship between the SFR derived from our calibration given by equation~\ref{sfr10my} and stellar masses for our sample.  The relation for the total masses shown by the red points and red line  is given by
\begin{equation}
  \log({\rm SFR}) = -8.01\pm0.09 + 0.93\pm0.01 \times \log(M_{total}) 
\end{equation}with an rms=0.313.  For comparison, we plot the relation derived from other commonly used samples of  star-forming galaxies in the literature. The green and cyan lines are from \cite{chang2015}. The latter represents their calibration using the values from \cite{brinch2004} for stellar masses and SFRs.  The orange line is  from \cite{elbaz2007} for their sample of blue star forming galaxies.  Even considering the scatter in these relations it is clear that the main sequence of HII galaxies (red points) is significantly steeper and stronger than the relation for more general samples of star-forming galaxies.  

The blue points in  the figure show the SFR for the starburst component alone. A linear fit (blue line) to this relation gives
\begin{equation}
 \log({\rm SFR}) = -6.24\pm0.06 + 0.92\pm0.01 \times \log(M_{young}) 
\end{equation}with an rms=0.297.
This relation can be interpreted as the empirically derived main sequence for single starbursts and represents the maximum star formation rates for starbursts. 

To illustrate this point Figure~\ref{ms_gals} (right panel) shows the relation between the so-called specific star formation rate (sSFR; star formation rate per unit mass) as a function of mass. As in the previous figure, the blue points represent the star formation per unit mass of stars formed in the current star formation episode ($M_{young}$) and the red points represent the current star formation rate over all stars ($M_{total}$) formed in the past history of the galaxy.  HII Galaxies fall well above the overall average for star-forming galaxies of log(sSFR) $\sim$ -10 yr$^{-1}$ \citep{guo2015}.

Thus, averaged over their entire lifetimes HII Galaxies have been forming stars at levels that approach the largest starburst galaxies such as ULIRGS represented in the figure by Arp 220 with log(sSFR) $\sim$ -8 yr$^{-1}$. However, if one considers the  specific star formation rate of the present burst alone (SFR per unit {\it young} stellar mass),  the current starbursts in HII Galaxies are producing new stars at a much higher rate of log(sSFR) $\sim$ -7 to  -6.5 yr$^{-1}$. By the same token, the present-day sSFR of ULIRGS are close to the upper limit set by HII galaxies when normalised by their young stellar masses.
 
Figure~\ref{ms_gals} also shows the SFR (left panel) and sSFR (right panel) for the prototypical Giant HII Region 30 Doradus in the LMC. \cite{doran2013}  derived a mass of $1.1\times10^5$\msun\ and a SFR of $0.073 \pm 0.04$ \msun$yr^{-1}$ for a Kroupa IMF from their stellar Lyman continuum census.  Although star formation in 30 Dor has spread over 5 Myr \citep{selman1999}, this is the closest example of a genuine real-life simple young massive  stellar population, and sets an upper limit to how fast star formation may occur in starbursts.  
The purple dashed-line in Figure~\ref{ms_gals} (right panel) shows the "speed limit" for star formation that is set by 30Dor. % which corresponds to its average age.  

As discussed above, simple dynamical arguments imply that objects substantially more massive than 30 Doradus, which is the case for all the HII Galaxies in our sample, cannot form stars significantly faster than 30Dor. This also explains the declining tilt of the sSFR of our galaxies shown in the figure: only low-mass HII Galaxies can harbour the most intense starbursts.
\subsection{The \lsigma\ relation}

The strong relation between \lhbeta\ and \myou\ found in this paper confirms that the \lsigma\ relation is indeed a correlation between the mass of the starburst component and the turbulence of the ionized gas. However, the relation remains empirical because we do not yet fully understand the origin of the gas turbulence in HII Galaxies. It could be due to gravity, if the gas clouds are virialized, or to the direct injection of mechanical energy from massive stars via stellar winds, or a combination of both.

  We used the $f_r$ factors derived from our SED fitting to correct the luminosities of the galaxies used in
\cite{mel2017} to study the scatter of the relation and found that the corrections actually increase the scatter quite substantially. This is probably due to the fact that the corrections expose the effect of a second parameter -- possibly the effective radius of the young component ($R$) as suggested by \cite{chavez2016} and expected if the gas is virialized. Unfortunately, however, good measurements of the effective radii 
of these galaxies are not yet available to verify, for example, whether $R\sigma^2$ correlates with \myou\ as expected if the gas is in Virial equilibrium with the stellar potential. 

\section{Concluding remarks}\label{conclusions}

We have studied a representative sample of the youngest \mbox{($<{\rm EW(H}\beta)>=50$\AA)}; highest excitation  and lowest metallicity (<12+logO/H>=8.2) HII Galaxies in the local universe ($z<0.4$) and find that, as a class, they have the following properties:

\begin{enumerate}

\item The correction factor $(1 +f_r)$ is typically between 1.5 and 2.5. This factor is derived from  our SED fitting procedure and is then applied on the observed ${\rm EW(H}\beta)$ in order to correct for the contribution of the underlying old stellar continuum and recover the true ${\rm EW(H}\beta)_{young}$.

 %The correction factor $(1 +f_r)$, derived from our SED fitting procedure, to be applied on the observed ${\rm EW(H}\beta)$ to correct for the contribution of the underlying old stellar continuum in order to recover the true ${\rm EW(H}\beta)_{young}$ is typically between 1.5 and 2.5.
  
  \item The star formation histories of HII Galaxies can be reproduced remarkably well by three bursts of star formation: (a) the current young burst, a few Myr old, that dominates the luminosity at all wavelengths but contains only a few percent of the total mass; (b) an intermediate age burst of a few hundred Myr; (c) and old stellar component (10 Gyr),  which together contain most of the mass.  Therefore, the past SF history is far more important in producing the bulk of the stellar mass in HII galaxies.
  
  \item At a given age, the \hbeta\ luminosity of HII Galaxies depends only on the mass of young stars. This implies that the  IMF of the ionizing clusters must be a universal function at least for massive stars, and  that only a relatively small fraction of Lyman continuum photons escape from the nebulae (case B photoionization). 
  
  \item The "main sequence" of star formation for HII Galaxies is significantly steeper and stronger than that of more massive star forming galaxies from the literature, while the present-day star formation rates of HII Galaxies are on average a factor of three larger than predicted by the H$\alpha$ Kennicutt relation.  Therefore, extreme care must be exercised when combining starburst galaxies with more normal galaxies to construct the overall "main sequence" of star-forming galaxies.

 \end{enumerate}

\section*{Acknowledgments}
We are grateful to  Denis Burgarella, the father of CIGALE, and to M\'ed\'eric Boquien for guiding us through our first steps with the code and answering numerous questions. M\'ed\'eric kindly wrote the special module to fit three stellar populations that we used in this work.  ET thanks Roderik Overzier for fruitful discussions, and Elena Terlevich for comments on the manuscript. JM acknowledges support from a CNPq {\it Ciencia Sem Fronteiras} grant at the Observatorio Nacional in Rio de Janeiro, and the hospitality of ON as a PVE visitor.  Finally, we thank the anonymous referee for his/her comments and suggestions to improve the paper.

Funding for the SDSS and SDSS-II has been provided by the Alfred P. Sloan Foundation, the Participating Institutions, the National Science Foundation, the U.S. Department of Energy, the National Aeronautics and Space Administration, the Japanese Monbukagakusho, the Max Planck Society, and the Higher Education Funding Council for England. The SDSS Web Site is http://www.sdss.org/.

The SDSS is managed by the Astrophysical Research Consortium for the Participating Institutions. The Participating Institutions 
are the American Museum of Natural History, Astrophysical Institute Potsdam, University of Basel, University of Cambridge, 
Case Western Reserve University, University of Chicago, Drexel University, Fermilab, the Institute for Advanced Study, the 
Japan Participation Group, Johns Hopkins University, the Joint Institute for Nuclear Astrophysics, the Kavli Institute for Particle
Astrophysics and Cosmology, the Korean Scientist Group, the Chinese Academy of Sciences (LAMOST), Los Alamos National
Laboratory, the Max-Planck-Institute for Astronomy (MPIA), the Max-Planck-Institute for Astrophysics (MPA), New Mexico State 
University, Ohio State University, University of Pittsburgh, University of Portsmouth, Princeton University, the United States 
Naval Observatory, and the University of Washington.

The entire GALEX Team gratefully acknowledges NASA's support for construction, operation, and science 
analysis for the GALEX mission, developed in corporation with the Centre National d'Etudes Spatiales of 
France and the Korean Ministry of Science and Technology. We acknowledge the dedicated team of engineers,
technicians, and administrative staff from JPL/Caltech, Orbital Sciences Corporation, University of California, 
Berkeley, Laboratoire d'Astrophysique Marseille, and the other institutions who made this mission possible. 
 
The UKIDSS project is defined in Lawrence et al. (2007). UKIDSS uses the UKIRT Wide 
Field Camera WFCAM (Casali et al. 2007).  The photometric system is described in 
Hewett et al. (2006), and the calibration is described in Hodgkin et al. (2008). The 
pipeline processing and science archive are described in Irwin et al. (2009, in prep) 
and Hambly et al (2008).

This publication makes use of data products from the Wide-field Infrared Survey Explorer, which is a joint project of the University 
of California, Los Angeles, and the Jet Propulsion Laboratory/California Institute of Technology, funded by the National 
Aeronautics and Space Administration

\bibpunct{(}{)}{;}{a}{}{,} % to follow the A&A style

\bibliographystyle{aa} % style aa.bst
\bibliography{cigale} % your references

\end{document}